\documentclass[preprint]{elsarticle}

\usepackage{hyperref}

\usepackage{amsmath}
\usepackage{cancel}

\usepackage[ruled,vlined]{algorithm2e}

\usepackage{calc}
\usepackage{tikz}

\usepackage{graphicx}
\usepackage{xcolor}
\definecolor{cc}{rgb}{1,0.5,0}
\definecolor{c2}{rgb}{0.4,0.64,0.66}
\usepackage{dcolumn}
\newcolumntype{d}[1]{D{.}{.}{#1}}

\usepackage{multirow}
\usepackage{caption}
\usepackage{subcaption}

\journal{Journal of Systems and Software}

\begin{document}

\begin{frontmatter}

\title{An Estimation of Distribution Algorithm based on interactions between requirements to solve the bi-objective Next Release Problem}

%\title{Elsevier \LaTeX\ template\tnoteref{mytitlenote}}
%\tnotetext[mytitlenote]{Fully documented templates are available in the elsarticle package on \href{http://www.ctan.org/tex-archive/macros/latex/contrib/elsarticle}{CTAN}.}

%% Group authors per affiliation:
\author[mainaddress]{Jos\'e del Sagrado}

\author[mainaddress]{Jos\'e Antonio Sierra Iba\~nez}

\author[mainaddress]{Isabel M. del \'Aguila}

\address[mainaddress]{Department of Informatics\\ University of Almer\'{i}a \\ Crtra. Sacramento S/N \\ 04120 La Ca\~nada (Almer\'{i}a Spain)}

\begin{abstract}
Selecting the appropriate requirements to develop in the next release of an open market software product under evolution, is a compulsory step  of each software development project.
This selection should be done by maximizing stakeholders' satisfaction and minimizing development costs, while keeping constraints. In this work we investigate what is the requirements interactions impact when searching for solutions of the bi-objective Next Release Problem. In one hand, these interactions are explicitly included in two algorithms: a branch and bound algorithm and an estimation of  distribution  algorithm (EDA). And on the other, we study the performance of these not previously used solving approaches by applying them in several instances of small, medium and large size data sets.
We find that interactions inclusion do enhance the search and when time restrictions exists, as in the case of the bi-objective Next Release Problem, EDAs have proven to be stable and reliable locating a large number of solutions on the reference Pareto front.
\end{abstract}


\end{frontmatter}

%\linenumbers

\section{Motivation}

In a software development project, stakeholders propose some desired functionalities that have to be filtered in order to define the set of features to include in the next software version. This decision task is still a challenging problem because of the complexity and the so different interactions between the  involved elements, such as the high number of stakeholders, the variety and heterogeneous nature of the variables to be reviewed or the uncertainty of the data used to solve it \citep{Ruhe2010}. This problem has been addressed by various authors paying attention to different aspects of release decisions, both outside the software development industry, such as those related to finding the best combination of features to include in a release sequence (i.e., release schedule) \cite{etgar2017,greer2004}, and within it, such as those that focus on defining the optimal set of features just for the next product release \cite{durillo2011,chaves2015a,Dominguez2019}.

Stakeholders' proposals are specified in terms of requirements, that are the conditions or capabilities required by a stakeholder to solve a problem or achieve an objective \cite{brennan2009}. Nonetheless, conflicts can arise because of the involvement of multiple stakeholders, requiring negotiation and agreements about the project scope. Therefore, requirements selection is pivotal because not all stakeholders' requirements can be met within the available time and the resource constraints.

Due to the computational complexity of the problem, it has been  formulated as an optimization problem with the aim to move software engineering problems from human-based search to machine-based search, using a variety of techniques from metaheuristic search, operations research and evolutionary computation paradigms \cite{harman2010}. This approach has been applied, successfully and prolifically, to different aspects of requirements selection such as requirements prioritization, requirements selection, release planning, Next Release Problem and  requirement triage \cite{pitangueira2015}. The applied solving strategy varies from  algae algorithms~\cite{piroz2021}, whale and grey wolf optimization~\cite{ghasemi2021}, bee colony approaches~\cite{alrezaamiri2020}, anytime algorithms~\cite{Dominguez2019}, clustering~\cite{sagrado2021,etgar2019}.

Requirements can not be studied in isolation, they are interrelated and affect each other in complex ways \cite{carls2001}. These interactions or dependencies make some requirements to precede or exclude others when are considered for their inclusion. So that interactions are also constraints on the problem. Within the domain of requirements selection, precedence has been represented as a graph \citep{ngo2004}, which was extended by including other types of dependencies between requirements \citep{sagrado2015}. Thereby, functional interactions are problem restrictions, being the precedence ones those that limit the search space \citep{sagrado2011}. 

In software engineering, the problem of finding an optimal set of requirements that maximizes the stakeholders' satisfaction, while minimizing the development effort, which is limited by the capacity of the development team, is known as the bi-objective next release problem (NRP) \cite{zhang2007, durillo2011}. This set will define the features included in the next version of the software product under development. Although many methods \cite{durillo2011, chaves2015a, sagrado2015, alrezaamiri2020, sagrado2021, ghasemi2021} have been employed to obtain a set of optimal solutions for this problem, dependencies are usually taken into account to rule out the solutions when they are evaluated, but not when requirements are assembled for constructing them. Our purpose is to investigate the impact that requirements interactions have in the search. For it, we will incorporate them into two algorithms  (a branch and bound algorithm and an estimation of distribution algorithm) for solving the bi-objective Next Release Problem. 

Estimation of distribution algorithms (EDAs) belong to the class of evolutionary algorithms. At each iteration they proceed by learning and sampling the probability distribution of the best individuals of the population \cite{Larranaga:2002}. Thus, the probabilistic model learned is in charge of capturing the relationships between the problem variables. Consequently, at each iteration most of the EDA effort is focused on learning this probabilistic model. Therefore, it is worth exploring whether the innate relationships between requirements, defined by their interactions, could alleviate the EDA bottleneck and be useful for improving the selection of requirements when applying these types of approaches. Specifically, we seek to answer the following questions:

\begin{itemize}
    \item  \emph{RQ1} How can requirements interactions be included in search algorithms to find the Pareto front of a next release problem?
    \item \emph{RQ2} Is there any improvement in incorporating the interactions between the requirements in the algorithms?
    \item \emph{RQ3} If a set of non-dominated solutions is required in a short time, do estimation of distribution algorithms find a high quality (i.e. well-spread) one?
\end{itemize}

The paper is organized as follows. In Section 2, we define concepts related to the software requirement selection problem and the formulation of the bi-objective Next Release Problem is also presented here. Interaction based search is the answer to research question 1 and is the topic of Section 3. A branch and bound algorithm and an estimation of distribution algorithm are described to solve the bi-objective Next Release Problem. Section 4 presents the analysis of the algorithms and the computational results, answering research questions 2 and 3. The identified threats to validity are reported in Section 5. The last section presents the conclusions. 

\section{Software requirement selection problem}

A necessary step at the beginning of each software project iteration is to select the appropriate requirements to achieve maximized business profit within a given resource bound, that is based on developers effort estimations and team capacity. Each client, considered according the market policies defined by the company, requests a fraction of the candidate's requirements, and for each one of them indicates the satisfaction they would obtain if it was included in the software product under construction. Not all the requirements can be satisfied, mainly due to the limited availability of resources and the existing interactions between them.

While requirements are present in every kind of software project, not in all  the  clients have the chance to include or evaluate their own requests to the development team, nor have the opportunity to be involved in the requirements selection tasks. Nonetheless, some other projects, such those that apply an agile development strategy or those that are open source, cannot go without clients involvement. The goal of new versions have mandatory be defined based on the clients suggestions, being agreed with them, as well.

The term Next Release Problem (NRP) was introduced by Bagnall et al. \cite{BAGNALL2001883}. In its original formulation, the objective is to select a subset of requirements which gives rise to a subset of clients to satisfy, in such a way that it maximizes the clients' satisfaction and the total estimated cost of this subset of requirements does not exceed a given budget. Consequently, this task could be understood as a single-goal optimization problem subject to a budget limit constraint. Such a combinatorial problem is an instance of the  knapsack problem that is $\mathcal{NP}$-hard \cite{almeida18}.

This mono-objective formulation to NRP was extended to a multi-objective one \cite{zhang2007}, which proposes the use of a bi-objective strategy where the problem is to select the set of requirements that maximize total satisfaction and minimize required cost (i.e. resources needed) meeting the constraints defined by the resource limits and the interactions. Thus, the solving algorithms return a set of solutions which are all efficient in the Pareto sense and can help software developers, when they face contradictory goals, to make a decision on which is the subset of requirements that has to be considered in the next development stages. Usually a commitment is required, because high satisfying subsets of requirements could not cover the full team capacity being good alternatives according clients and developers for a given project iteration. Different meta-heuristic optimization methods have been used to obtain the set of solutions for the bi-objective NRP, such as integer linear programming \cite{veerapen2015}, anytime algorithms \cite{Dominguez2019}, multi-objective teaching learning based optimization \cite{chaves2015b}, genetic algorithms \cite{zhang2007, durillo2011}, swarm intelligence \cite{chaves2015a}, ant-colony optimization \cite{Jiang2010,sagrado2015}, bee-colony optimization \cite{alrezaamiri2020} and clustering algorithms \cite{sagrado2021, HUJAINAH2021}, among others, are only some examples.

Because of interactions, it makes sense to explore only the areas of the NRP search space containing valid solutions, which will also translate into an improvement in the search process \cite{sagrado2011}. When searching for solutions, validity can be ensured in advance by inserting requirements in the solution in an order that satisfies the interactions, instead of inserting requirements randomly and then checking the validity of the solution. Precedence (some requirements should be done before others could be included) naturally leads to a disposition of requirements that gives support to this search strategy.

\subsection{Problem definition}

Let \(\mathbf{R} = \{r_1,r_2,\ldots,r_n\}\) be the set of requirement to be considered. These represent new functionalities to the current system suggested by a set of $m$ clients and are the candidates for to be joined in the next software release. The revenue or satisfaction  estimated for each requirement is calculated by the weighted sum of individual clients' revenues according their importance for the project. It is represented by the set of satisfactions, $\mathbf{S} = \{s_1, s_2, \ldots, s_n\}$. Besides, each $r_j$ has an associated cost, $e_j$, that estimates the development effort needed for its implementation, $\mathbf{E}= \{e_1, e_2, \ldots, e_n\}$. Thus, given a subset of requirements $\mathrm{\hat{\textbf{R}}}$ included in $\mathrm{\textbf{R}}$ its satisfaction and effort can be can be defined, respectively, as:

\begin{align}
    \mathrm{sat}(\mathrm{\hat{\textbf{R}}})= \sum_{r_j \in \mathrm{\hat{\textbf{R}}}} s_j
\end{align}

\begin{align}
    \mathrm{eff}(\mathrm{\hat{\textbf{R}}})= \sum_{r_j \in \mathrm{\hat{\textbf{R}}}} e_j
\end{align}

In every software release, there are certain restrictions that must be met. There is a cost limit $B$, representing the amount of available resources, that cannot be overrun. In addition, requirements interactions have also to be considered and preserved. There are two main groups of requirement interactions: \emph{functional} and \emph{value based} \cite{sagrado2015}. The interactions in the \emph{functional interactions} group are 
\begin{itemize}
    \item \emph{Implication or precedence}, $r_i \Rightarrow r_j$, indicates that requirement $r_i$ has to be included (implemented) in a release before requirement $r_j$.
    \item \emph{Combination}, $r_i \Leftrightarrow r_j$, means that both requirements $r_i$ and $r_j$ have to be included together in a release.
    \item \emph{Exclusion}, $r_i \cancel{-} r_j$, implies that both requirements $r_i$ and $r_j$ are incompatible and can not be included in a release.
\end{itemize}
While the \emph{value-based} interactions group includes both \textit{revenue-based} and \textit{cost-based} interactions. The former involve changes in profit and the latter changes in the amount of resources needed to implement each requirement. Usually, in the bi-objective NRP, only functional interactions are considered, and this is the approach we shall follow from now on.

To solve a bi-objective NRP, a subset of requirements $\mathrm{\hat{\textbf{R}}}$ included in $\mathrm{\textbf{R}}$, which maximize satisfaction and minimize development effort has to be selected while cost limit and requirements interactions constraints are preserved. Therefore, the requirements selection problem for the next software release can be formulated as the next bi-objective optimization problem:

\begin{equation}
    \begin{aligned}
    & \max_{{\hat{\mathbf{R}}} \subseteq {\mathbf{R}}} && \mathrm{sat}(\hat{\mathbf{R}}) \\
    & \min_{{\hat{\mathbf{R}}} \subseteq {\mathbf{R}}} && \mathrm{eff}(\hat{\mathbf{R}}) \\
    & \textrm{s.t.} && \mathrm{eff}(\hat{\mathbf{R}}) \leq B \\
    %& \textrm{s.t.} && \sum_{r_j \in \hat{\textbf{R}}} e_j \leq B \\
    &               && \hat{\mathbf{R}} \textrm{ fulfills all functional dependencies}\\
    \end{aligned}
\end{equation}

\subsection{Multi-objective optimization background}

A multiobjective optimization problem can be defined as minimizing (or maximizing) simultaneously multiple objective functions $F(\vec{x}) = (f_1(\vec{x}), \ldots , f_k(\vec{x}))$
\begin{equation} \label{eq:genop}
    \begin{array}{ll}
         \min (F(\vec{x}) = & (f_1(\vec{x}), \ldots, f_z(\vec{x})))  \\
         \text{subject to} & \vec{x} \in \mathbf{X}, \\
                           & h_1(\vec{x}) \leq g_1, \ldots, h_l(\vec{x}) \leq g_l
    \end{array}
\end{equation}

\noindent where $\vec{x} \in \mathbf{X}$ is a feasible solution, $\mathbf{X}$ is the feasible set of vectors, typically $\mathbf{X} \subseteq \mathbf{R}^n$, $h_1(\vec{x}) \leq g_1, \ldots, h_l(\vec{x}) \leq g_l$ are some constraint functions that define the feasible set and $g_1, \ldots, g_l$ are some constant values~\cite{coello2007}. 

Let  $\vec{u}, \vec{v} \in \mathbf{X}$, $\vec{u} \neq \vec{v}$, be two distinct feasible solutions, we say that $\vec{u}$ \emph{dominates} (i.e. is preferred to) $\vec{v}$, $\vec{u} \succ \vec{v}$ if and only if $\forall i \in \{1, \cdots, z\}$, $f_i(\vec{u}) \leq f_i(\vec{v})$ and $\exists j \in \{1, \cdots, z\}$, $f_j(\vec{u}) < f_j(\vec{v})$. A solution $\vec{u}$ is said to be \emph{Pareto-optimal} if and only if there is no other vector $\vec{v}$ that dominates it. The \emph{Pareto front} is the set of all Pareto-optimal solutions. A simple way to obtain this set is to check whether each of the feasible solutions $\vec{u} \in \mathbf{X}$ is \emph{dominated} or not. If it is not, it is added to the \emph{Pareto front}. 

According to this definition, an \ensuremath{\mathsf{NRP}} can be formulated as a multiobjective optimization problem, where the feasible solutions are sets of requirements and the objective and constraint functions (i.e. satisfaction and effort) are related to these sets. Objectives and constraints may be linear or nonlinear and continuous or discrete in nature based on the available information about requirements.

In the context of bi-objective NRP, $\vec{u}$ and $\vec{v}$ are sets of requirements, \emph{sat} (satisfaction) and \emph{eff} (effort) are the constraint functions and we say that $\vec{u}$ \emph{dominates} $\vec{v}$ if and only if $\textit{eff}(\vec{u}) \leq \textit{eff}(\vec{v})$, $\textit{sat}(\vec{u}) > \textit{sat}(\vec{v})$. Thus, Pareto dominance is used to establish both the set of Pareto optimal solutions (i.e. the candidate groups of requirements to be implemented in the next release) and their projection into the objective space, that is, the Pareto front.

\subsection{Interaction graph definition}
\label{subsec:interactions_transformation}

Functional interactions can be explicitly represented as a graph \(G = (\mathbf{R}, \mathbf{I}, \mathbf{J}, \mathbf{X})\) where:
\begin{itemize}
    \item \textbf{R} (requirements set) is the set of nodes.
    \item \textbf{I} is the set of implication dependencies. Each pair $(r_i,r_j) \in \mathbf{I}$ corresponds to an implication interaction and will be represented as a directed link $r_i \rightarrow r_j$, dictating the implementation precedence between requirements. 
    \item \textbf{J} is the set of combination dependencies. Each pair $(r_i,r_j) \in \mathbf{J}$ corresponds to a combination interaction and will be represented as a bidirectional link $r_i \leftrightarrow r_j$, indicating that both requirements have to be included in the same release.
    \item \textbf{X} is the set of exclusion dependencies. Each pair $(r_i,r_j) \in \mathbf{X}$ corresponds to an exclusion interaction and will be represented as a crossed undirected link $r_i \cancel{-} r_j$, meaning that both requirements are incompatible and can not be included in the same release.
\end{itemize}

\begin{figure}[htbp]
   \centering
   \includegraphics{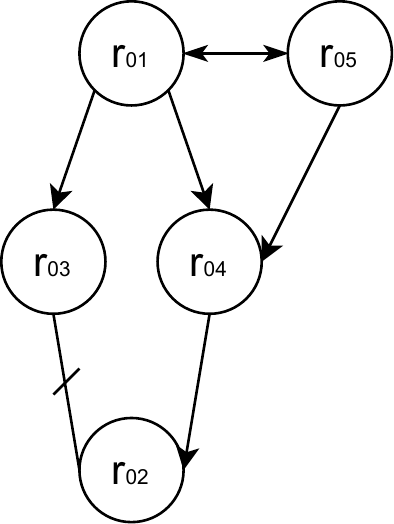}
   \caption{Interactions graph}
   \label{fig:grafo_dependencias}
\end{figure}

For example, Figure \ref{fig:grafo_dependencias} shows the graph representing the set of requirements $\textbf{R}=\{r_{01}, \cdots, r_{05}\}$ and the following sets of functional interactions: \mbox{$\textbf{I}=\{(r_{01}, r_{03}), (r_{01}, r_{04}), (r_{04}, r_{02})\}$}, \mbox{$\textbf{J}=\{(r_{01}, r_{05})\}$} and \mbox{$\textbf{X}=\{(r_{03}, r_{02})\}$}.

Implication interactions lead to a requirements' organization as they indicate precedence between them. So, a searching algorithm could proceed by inserting requirements in an order such that requirements interactions are satisfied, which would also ensure in advance the validity of the solution. Since combination and exclusion interactions must also be considered, the interaction graph has to be transformed so that only implication interactions prevail as a guide to find valid solutions.

\begin{figure}[ht]
    \centering
    \fontsize{9}{12}\selectfont
    \includegraphics{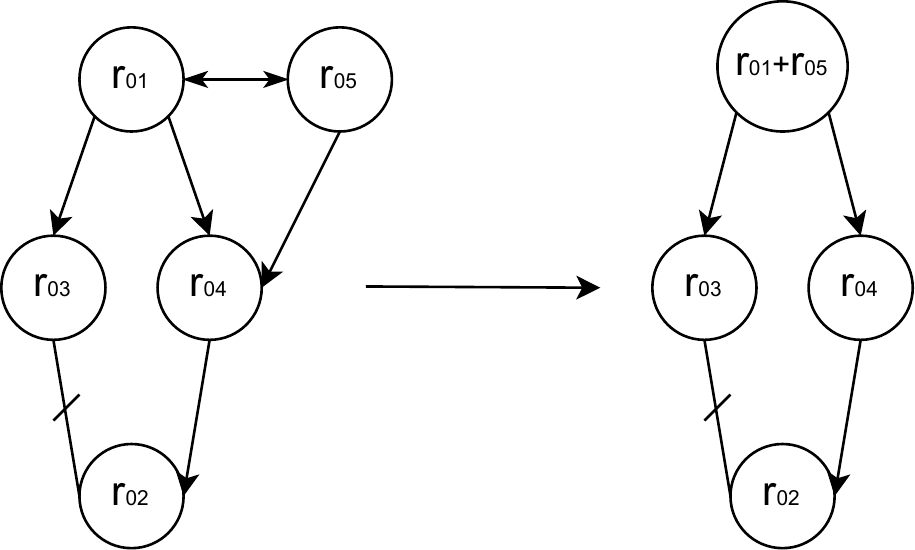}
    \caption{Transformation of combination interactions.}
    \label{fig:grafo_combinacion}
\end{figure}

First, \emph{combination interactions} are made over. Each pair $(r_i,r_j) \in J$ will be transformed into a new requirement $r_{i+j}$ whose satisfaction, $s_{i+j}=s_i+s_j$, and effort, $e_{i+j}=e_i+e_j$, values will be the sum of the satisfaction or the effort of the requirements involved, respectively. Occurrences of $r_i$ and $r_j$ in exclusion or implication interactions are replaced by the new requirement $r_{i+j}$, avoiding the repetition of pairs. As result of this step, a new functional interaction graph $G^{\prime} = (\mathbf{R}', \mathbf{I}, \mathbf{J}, \mathbf{X})$ in which, $\mathrm{\textbf{J}}=\emptyset$, combination interactions are deleted is obtained. Figure \ref{fig:grafo_combinacion} shows the graph  after applying this transformation for the  example with 5 requirements.

Next, exclusion interactions are transformed. This transformation involve the inclusion of additional nodes in the graph. Let us consider the exclusion  $(r_i, r_j) \in \mathbf{X}$. It is replaced by adding two variables $I_{r_i}$ and $I_{r_j}$, and two directed links $(I_{r_i}, r_j)$ and $(I_{r_j}, r_i)$. The purpose of an indicator variable, such as $I_{r_i}$, is to represent if it is possible to include the other requirement $r_j$ in a solution $\mathbf{R}_s$ without violating the given exclusion interaction with $r_i$, and is defined as

\begin{align} \label{eq:indicadora}
    I_{r_i} = \begin{cases}
    			0, & \textup{if } r_i \in \mathbf{R}_s \textup{ and }(r_i, r_j) \in \mathbf{X},\\
    			1, & \textup{otherwise. } 
    			\end{cases}
\end{align}

That is, $r_j$ can be included in a solution $\mathbf{R}_s$ without violating exclusion when $I_{r_i} = 1$, $r_i \notin \mathbf{R}_s$ and $pa(r_j) \in \mathbf{R}_s$, where $pa(r_j)$ is set of requirements that are the origin of a directed link that ends in $r_j$ in the graph.

\begin{figure}[ht]
   \centering
   \fontsize{9}{12}\selectfont
   \includegraphics{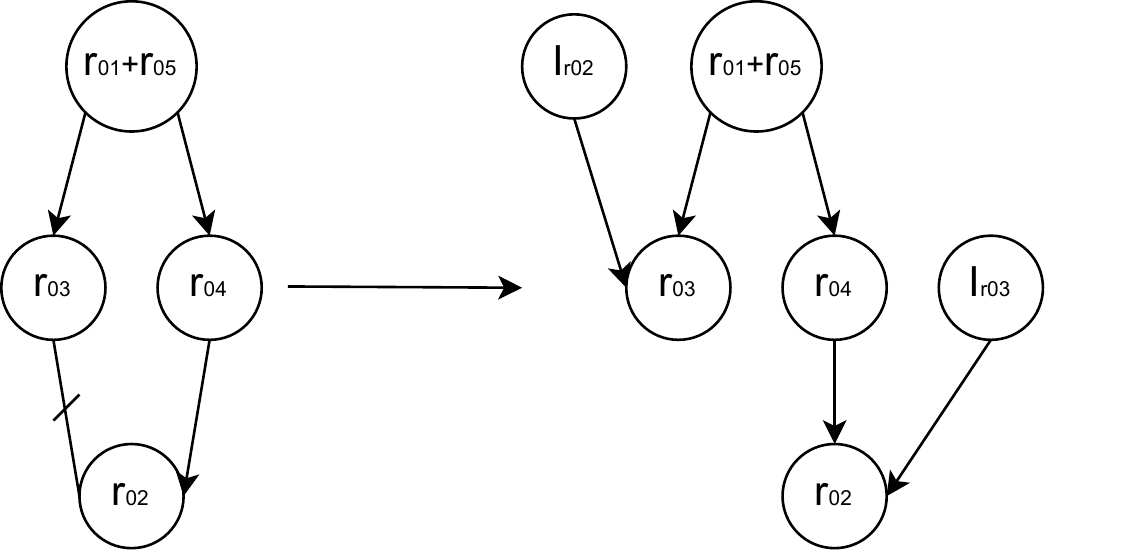}
   \caption{Transformation of exclusion interactions.}
   \label{fig:grafo_exclusion}
\end{figure}

The final interaction graph, $G''=(\mathbf{R}'\cup\mathbf{V}_I,\mathbf{I},\mathbf{J},\mathbf{X})$, obtained after transforming combination and exclusion, contains only implication dependencies, therefore $\mathrm{\textbf{J}}=\emptyset$ and $\mathrm{\textbf{X}}=\emptyset$. Figure \ref{fig:grafo_exclusion}
shows the final graph obtained after processing the exclusion interaction in the example with 5 requirements.

When searching for solutions of the NRP, requirements interactions have to be preserved. Let's consider the stage where the requirements are assembled to build a valid solution. A requirement $ r_i $ can be added if every requirement on any directed path in the interaction graph $G''$ that goes from the root to itself is contained in the solution. Hence, requirements can be ordered into a sequence by making a requirement $r_i$ to come after all of its predecessors in the interaction graph. A sequence obtained in this way from a directed acyclic graph is called an ancestral (or topological) ordering. The sequences $\sigma_1 = \{ r_{01}+r_{05} \prec r_{03} \prec r_{04} \prec r_{02}\}$ and $\sigma_2 = \{r_{01}+r_{05} \prec r_{04} \prec r_{03} \prec r_{02}\}$ are examples of ancestral orderings in the final interaction graph for the problem with 5 requirements (see Figure \ref{fig:grafo_exclusion}).

Search algorithms can benefit from ancestral ordering and explore possible solutions by inserting requirements in that order. In this way they incorporate requirements interactions into the search process, and we get the answer to \textbf{RQ1}. The next stage is to validate how search algorithms can use them to guide their execution.

\section{Interaction-based search}

This section studies two alternatives to make use of the ancestral order defined by the interaction graph to solve a bi-objective NRP. A branch and bound exact algorithm and an estimation of distribution algorithm are described to explore the search space of a bi-objective NRP and find the set of non-dominated solutions (Pareto front). Besides, because of the evolutionary nature of EDA, some aspects about the population initialization and solutions replacement are also described.

\subsection{Branch and bound algorithm}
Branch and bound algorithms \cite{Morrison2016} are indicated for finding exact solutions to NP-hard optimization problems, like the bi-objective NRP. In this particular problem, they explore the search space by building a search tree, whose nodes will be sets of requirements. Instead of generating each combination of requirements and testing if it satisfies all the interactions to be a solution to the NRP problem (i.e. exhaustive search), we can start from an empty solution and try to add (or not) a requirement following an ancestral ordering (i.e. branch), discarding this partial solution as soon as any of the interactions are not fulfilled (i.e. bound). Algorithm \ref{alg:branch_bound} shows this process in detail. Note that \emph{pruning} is done in the \textit{if-sentence}, when the modified partial solution does not fulfil any of the NRP interactions.  

\vspace{6pt}

\begin{algorithm}[H]
\SetAlgoLined
\KwData{A bi-objective NRP and $\sigma$ an ancestral ordering of its requirements}
\KwResult{The set of solutions for the bi-objective NRP}
 
 Create an empty list of partial solutions\; 
 \For{each $r_i$ following the ancestral ordering $\sigma$}{
    Create an empty list of modified partial solutions\; 
    \While{the list of partial solutions is not empty}{
    Take a partial solution from the list\;
    Set $r_i = 0$ in the partial solution\;
    Add the modified partial solution to the list of modified partial solutions\;
    Set $r_i = 1$ in the partial solution\;
    \If{the modified partial solution fulfils NRP interactions}{
       Add the modified partial solution to the list of modified partial solutions\;
    }
    }
    The list of modified partial solutions becomes the list of partial solution list\;
 }
 Return the solutions in the partial solution list\;
 \caption{Branch and bound algorithm}
 \label{alg:branch_bound}
\end{algorithm}

\vspace{6pt}

\begin{figure}[ht]

  \includegraphics[width=0.9\columnwidth]{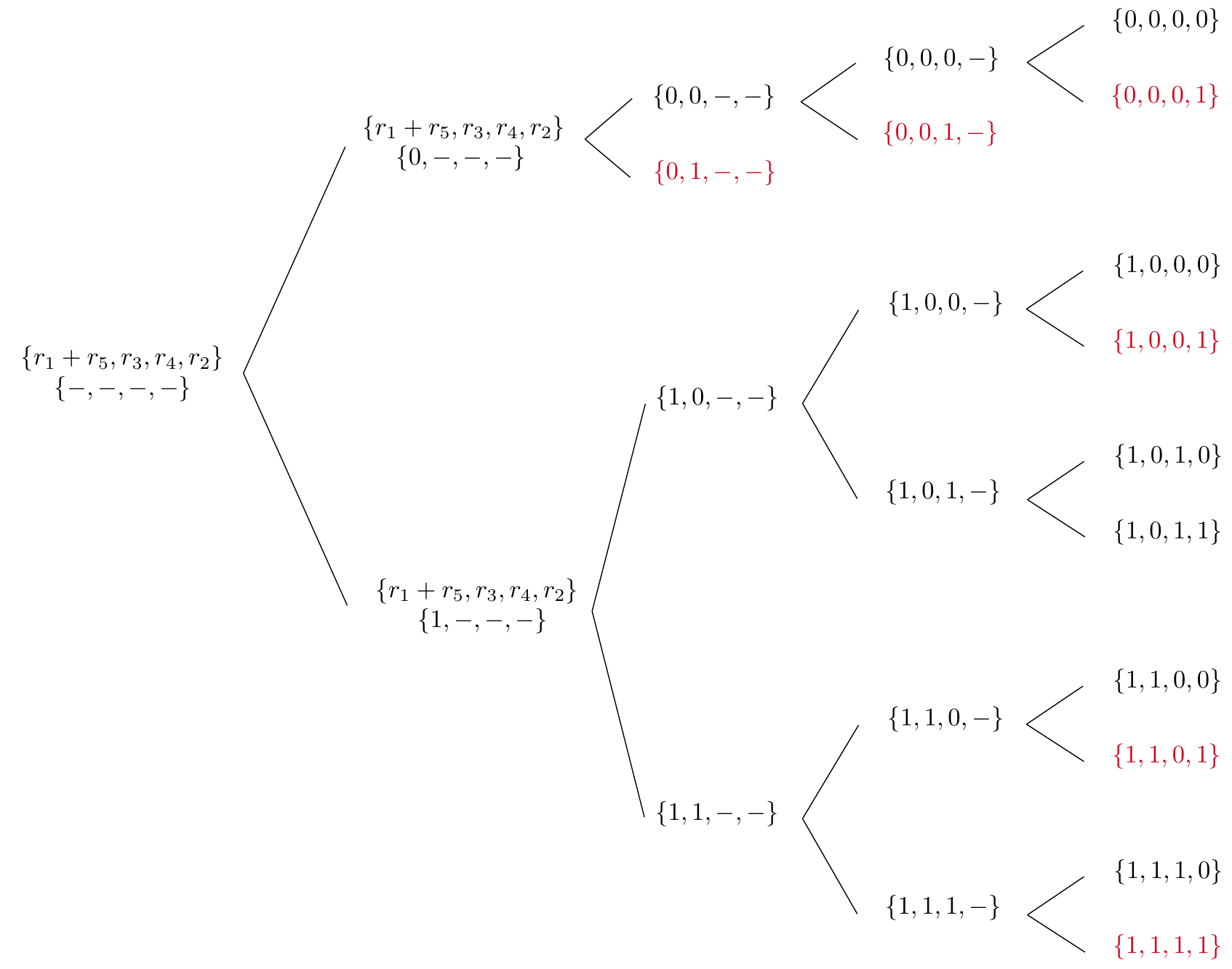}
   \caption{Branch and bound search tree. The requirement sets are represented between braces, 1 indicates that the requirement is included, 0 that it is not included and - that it is pending to be included or not.}
   \label{fig:BBSearchTree}
 \end{figure}

The algorithm performance depends on pruning. It will be more efficient if pruning is done at the beginning of a branch. Whilst, if no pruning is done, the algorithm will degenerate into an exhaustive search. Thus, the worst-case running time for the branch and bound algorithm \cite{Morrison2016} is $\mathit{O}(k \cdot b^d)$, where $b$ is the search tree branching factor, $d$ is the depth of this tree and $k$ is a bound on the time needed to explore a partial solution. Figure \ref{fig:BBSearchTree} illustrates how the brunch and bound algorithm explores the search space for the 5 requirements bi-objective NRP problem using $\sigma_1 = \{ r_{01}+r_{05} \prec r_{03} \prec r_{04} \prec r_{02}\}$ as ancestral ordering. Requirement sets depicted in red indicate that interactions are not fulfilled and, if possible, pruning is applied, avoiding the exploration of 8 requirement sets which means avoiding exploring 25.81\% of the search tree.

\subsection{Estimation of distribution algorithm}
Estimation of Distribution Algorithms (EDAs) belong to the class of evolutionary algorithms. In these algorithms, each solution is called \emph{individual} and a set of individuals forms a \emph{population} that evolves as the search progresses. The basic idea consists in inducing a probabilistic model from the best individuals in the population. Then, from the current population, EDAs build a probabilistic model and, by sampling it, obtain a new population. This search procedure is repeated until certain stopping criteria are satisfied. Algorithm \ref{alg:EDA} shows the basic steps of an EDA. A compilation and review on EDAs can be found in \cite{Larranaga:2002}.

\begin{algorithm}
\SetAlgoLined
\KwData{A bi-objective NRP} %and $\sigma$ an ancestral ordering of its requirements}
\KwResult{A set of solutions for the bi-objective NRP}
 
 $P_0 \gets$ Generate an initial population containing $m$ candidate solutions\;
 $M_0 \gets$ Learn a probabilistic model from $P_0$\;
 $j \gets 1$\;
 \While{ the stop criteria are not met}{
    $S_j \gets$ Sample $M_{j-1}$ and obtain $k \le m$ candidate solutions\;
    $P_j \gets$ Replace $S_j$ in $P_{j-1}$ according to the evaluation of individuals\;
    $M_j \gets$ Learn a probabilistic model from the population $P_j$\;
    $j \gets j + 1$\;
  } 
 Return the solutions in the last population $P_j$\;
 
 \caption{Estimation of distribution algorithm}
 \label{alg:EDA}
\end{algorithm}

An EDA explores the bi-objective NRP search space in two steps. First, from a set of selected candidate solutions, it learns a probabilistic graphical model that captures interactions between requirements. Next, it obtains a new set of solutions by sampling this model and replacing solutions based on their evaluation (i.e. satisfaction and effort). In this process, the main computational effort is placed in the learning step because a probabilistic graphical model has to be learnt every time a population is obtained. But, here is where interactions between requirements come into play, downsizing the execution resources needed by the EDA. 

Once the interaction graph has been constructed (see section \ref{subsec:interactions_transformation}) it does not change, remaining invariable, and will be established as the fixed structure of the probabilistic graphical models that indicates the factorization of the joint probability distribution. Therefore, the learning step in each iteration of the EDA is reduced to learn from a given population only the parameters associated to the directed acyclic graph that represents requirements interactions (i.e. learn the conditioned probability distributions of any requirement given its parents in this graph).

\subsubsection{Probabilistic graphical model for the NRP}
Formally, let $\mathbf{G} =(\mathbf{R}\cup\mathbf{V}_I,\mathbf{I},\mathbf{J},\mathbf{X})$ with $\mathrm{\textbf{J}}=\emptyset$ and $\mathrm{\textbf{X}}=\emptyset$, the directed acyclic graph obtained after processing combination and exclusion interactions in a NRP. The set of parents  $pa(x_i)$, associated to each variable $x_i \in \mathbf{R}\cup\mathbf{V}_I$, contains all variables that take part in a direct link of $G$ that ends in $x_i$. So, given a partial solution $\mathbf{R}_s$, a requirement $r_i$ can be added to it if all of its parents $r_j \in pa(r_i)$ are present in the candidate solution, $pa(r_i) \subseteq \mathbf{R}_s$. Otherwise, $r_i$ can not be included in $\mathbf{R}_s$. 
To complete the probabilistic graphical model, associated to each each requirement $r_i$ there is a conditional probability $p(r_i \vert  pa(r_i))$ which is defined as follow:

\begin{enumerate}
    \item If all $r_j \in pa(r_i)$ are present (i.e. $r_j = 1$) in the candidate solution $\mathbf{R}_s$, $pa(r_i) \subseteq \mathbf{R}_s$, then $r_i$ can be included (i.e. $r_i=1$) or not (i.e. $r_i=0$) in $\mathbf{R}_s$ with equal probability:
    \begin{equation}
       p(r_i \vert pa(r_i)) = \frac{1}{2}.
    \end{equation}
    That is to say, we can include $r_i=1$ in the candidate solution $\mathbf{R}_s$, because all implication interactions are fulfilled (all $r_j \in pa(r_i)$ are present in the candidate solution $r_j = 1, r_j \in \mathbf{R}_s$) and no exclusion interaction is active
    (all $I_k \in pa(r_i)$ are set to 1, $I_k = 1$). % \subseteq \mathbf{R}_s$.
    \item Otherwise, $r_i$ can not be included in $\mathbf{R}_s$ and
    \begin{equation}
        p(r_i=1 \vert pa(r_i)) = 0,
    \end{equation}
    because an implication interaction is violated (i.e. a requirement $r_j \in pa(r_i)$ is not present in the candidate solution, $r_j = 0, r_j \notin \mathbf{R}_s$) or an exclusion interaction is active (exists $I_k \in pa(r_i)$ which is set to 0, $I_k = 0$) or both.
\end{enumerate}
It is worth to note that the set of indicators variables $\mathbf{V}_I$ acts as an evidence set for exclusion interactions. Initially, when the candidate solution is empty, all variables in this set will be initialized to $1$. Then, if an exclusion interaction is activated  due to the presence of a requirement $r_k$ in the candidate solution, the value of an indicator variable $I_k$ will be changed to $0$ (see eq. \ref{eq:indicadora}). Therefore, the factorization of the joint probability distribution of the requirements given the evidence provided by the set of indicator variables can be written as:
\begin{equation}
    p(r_1, r_2, \cdots, r_n \vert V_I) = \prod_{i=1}^n p(r_i \vert pa(r_i)).
\end{equation}

\begin{figure}[ht]
  \centering
  %\fontsize{9}{12}\selectfont
  \includegraphics[width=0.8\columnwidth]{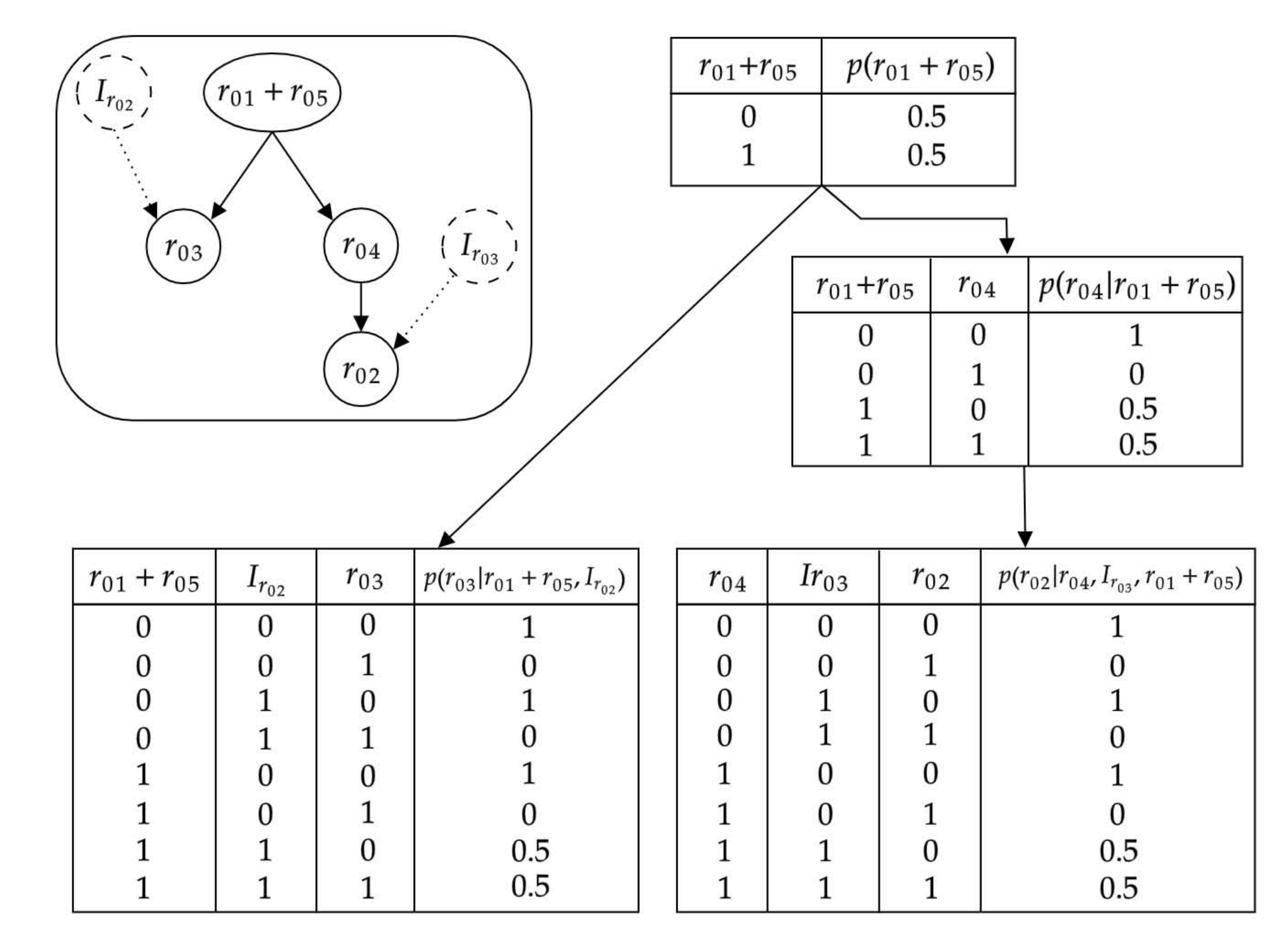}
  \caption{Graph after transforming interactions and the initial conditional probabilities.}
  \label{fig:initialBN}
\end{figure}

Figure \ref{fig:initialBN} shows the initial probabilistic graphical model defined for the 5 requirements NRP problem. Infeasible solutions, like $\{r_{01}+r_{05}, r_{03}, r_{02}\}$, have a factorized probability value, $p(r_{01}+r_{05} = 1, r_{03} = 1, r_{04} = 0, r_{02} = 1
\vert I_{r02}=0, I_{r03} = 0)$, equal to 0.

Once the initial probabilistic graphical model is defined, the EDA will generate a population by sampling it. Then, after replacement has been taken place, the conditional probabilities $p(r_i \vert pa(r_i))$, for the situation in which interactions are fulfilled and $r_i$ can be included in a candidate solution, will be estimated from the population (i.e. data) by using an $m$-estimator \cite{Cestnik:90}
\begin{equation}
    p(r_i \vert pa(r_i)) = \frac{N(r_i, pa(t_i)) + m \cdot p}{N(pa(r_i)) + m},
\end{equation}
where $N(pa(r_i))$ is the number of occurrences containing the values of the requirements indicated in the configuration of $pa(r_i)$, $N(r_i, pa(t_i))$ is the number of occurrences containing the values of the requirements indicated in the configurations of $r_i$ and $pa(r_i)$, $p = p(r_i)$ is the a priory probability of $r_i$, and $m$ is the \emph{equivalent data size}, a constant that can be freely chosen. The higher the value chosen for $m$, the greater the weight of the prior probability with respect to the value obtained from the frequency in the data.

\subsubsection{Sampling the probabilistic graphical model}
In each iteration of the the EDA, a new population is obtained by simulating the probability distribution encoded in the probabilistic graphical model learnt for the interaction graph of the NRP. Simulation methods can be stochastic or deterministic. Stochastic ones generate the sample from the joint probability distribution using random mechanisms, while deterministic methods generate the sample systematically. In this paper, we  use two simulation methods: \emph{Probabilistic Logic Sampling} (PLS) \cite{Henrion:88}, an stochastic method, and \emph{maximum probability search} method \cite{Poole:93}, a deterministic one. Both methods consider the variables following an ancestral ordering.  Other simulation methods can be found in \cite{Castillo:97}.

In \emph{Probabilistic Logic Sampling} \cite{Henrion:88}, each variable is instantiated forwards, that is, a variable can not be instantiated until all its parents have been instantiated. The method starts by generating a random number for each root variable of the network, and instantiates it by taking its a priori probability into account. Then, for each of the remaining variables, a value is simulated according the probability distribution $p(x_i \vert pa(x_i))$.

\emph{Maximum probability search} \cite{Poole:93} generates a tree following, at each moment, the branch with the highest probability. That is, given an ancestral ordering of the variables, it takes the first one and creates as many branches as possible values the variable can have, and chooses a branch with the highest probability. Then it takes the next variable in the order and increases the previously chosen branch with as many branches as the second variable can have. The probability of these branches is obtained by multiplying the probabilities (marginal or conditional) corresponding to the values that the variables have in each branch. This process continues until a complete realization of all the variables have been obtained. And so on, the search continues from the partial branch with maximum probability until enough complete realizations (i.e. possible solutions) have been found.

\subsubsection{Population initialization}
When initializing the population, we use three different alternatives to distribute the solutions in a way that covers as much as possible the full range of possible solutions. The first option is to generate solutions randomly over the range using a uniform distribution. Besides, it is possible to use the initial probabilistic graphical model, derived from the NRP, to generate this population by sampling. In this case, we can make use of probabilistic logic sampling and maximum probability search. Due to its nature, Probabilistic logic sample will return a more random population, whereas the population generated by maximum probability search will contain more promising solutions (those with maximum probability). Whatever the population initialization method used, the idea is to maintain a balance between the diversity of solutions and the time taken to obtain them.

\subsubsection{Solutions replacement}
Between two consecutive iterations the replacement (see algorithm \ref{alg:EDA}) is in charge of selecting, through a fitness-based process, individual solutions from the current population and the new set of solutions obtained by sampling. Therefore, to guarantee the solutions quality when constructing a new population the principles of diversity preservation and idiosyncrasy should be addressed. The best solutions (i.e. idiosyncrasy) from the current population and sample should be allowed to carry over to the next, unaltered, while at the same time less good solutions (i.e. diversity) should also be included to make possible population evolution and avoid the search to get trapped. 

Following these principles, the applied strategy starts by merging the current population and the obtained sample. Then, it proceeds to select the set of non-dominated solutions. %\color{magenta} 
If this set exceeds the population capacity, the new population is filled with the non-dominated solutions that achieve higher satisfaction. Otherwise, 
\color{black} the selected solution set is deleted from the merge and stored into the new population. This process of selection-deletion-storage is repeated until the capacity of the new population is filled up.

\section{Experiments}

The experiments let us to validate the research questions defined in this paper. By one hand, experimentation allows the validation of the model defined to incorporate interactions in the search process. That is, if the interaction graphs is useful enough for building the Pareto front in a bi-objective NRP (\textbf{RQ1}). On the other hand, the followed empirical protocol validates whether incorporating requirements interactions enhance the search process (\textbf{RQ2}) and evaluate the behaviour and performance of the EDAs, \textbf{RQ3}.
In the bi-objective NRP instance, we first use brute-force search (i.e. exhaustive search) to obtain the Pareto front, and then compare the time it takes with that of the branch-and-bound algorithm. With this direct comparison, we get a first insight on the impact interactions have on the search, \textbf{RQ2}. Then, we go an step further and apply the EDA. The approximate front it obtains and the time taken are compared with the exact front and the times set by the exact search algorithms, respectively. 

However, in medium and large (i.e. real) NRP instances the size of the search space is too large for the brute-force algorithm (and hence for the branch-and-bound algorithm). In this situation, 
we will divide the NRP instance ad-hoc into two sub-problems, apply exact (brute-force and branch-and-bound) algorithms to find the solutions to each sub-problem, and combine them to construct the Pareto front. Thus, the time spent on the construction of the exact front will correspond to the sum of the times spent on the search for solutions in each sub-problem and on their combination. In this way, we can compare the times taken by the exact algorithms to see the impact of interactions, and evaluate the behaviour and performance of EDAs as before. 

\subsection{Data sets}
We have used three data sets to test the effectiveness of our approach, comprising instances of software development problems of small, medium and large size that contain dependencies. The first one (\emph{NRP20}) is taken from \cite{greer2004}. It comprises $20$ requirements and $5$ clients both receiving an score from $1$ to $5$. Besides, each requirement has an associate effort scored between $1$ and $10$. Interactions between requirements include implication and combination interactions.
The second one (\emph{NRP50}) is taken from \cite{agarwal2014}, where the authors perform a case study of a project (planned in $2008$) derived from the well-known word-processing tool MS Word. The raw data has been pre-processed to match problem formulation. In the case at hand, it comprises $50$ requirements and $81$ functional interactions that have been elicited from $4$ clients. Each client gives a value to the requirements. The effort estimated in the development of each requirement has been measured in person-hours by the development team, considering all the stages in software development. The third data set (\emph{NRP100}) appears in \cite{sagrado2015} and includes $100$ requirements, $5$ clients and $44$ (implication and combination) requirements interactions. \color{magenta}It was generated following the patterns of agile software development methods. \color{black}
The development effort of a requirement is taken from real agile software projects developments and given in a $1$ to $20$ range. The maximum development effort is established in $20$ effort units, which can be translated into $4$ weeks (a usual time box in agile Software Engineering methods, e.g. Scrum proposes iterations between 2 and 4 weeks \cite{schwaber2001}). Clients make an assignment of the benefit of including a new requirement, using a coarse-grained scale. They simply place the requirements into one of three categories \cite{simmons2004, wohlin2012}: (1) inessential, (2) desirable, and (3) mandatory. \color{black} From these data sets, nine instances of NRP have been defined by fixing three different effort limits that correspond to 30\%, 50\% and 75\% of the total effort (i.e. the sum of efforts of the requirements in the data set).

\subsection{Pareto fronts comparison}
We use the hypervolume indicator \cite{Zitzler1998} as a measure of the objective space that is dominated by the points computed by the algorithms. It represents the union of the regions of all the rectangles that are dominated by the non dominated points. As a reference point to calculate the hypervolume we consider the Nadir point (B, 0), being B the NRP effort limit and 0 satisfaction. 

The hypervolume indicator, due to its properties, has been proposed as a guidance criterion for accepting solutions in Multiobjective Evolutionary Algorithms \cite{beume2009}. But, considering volume as the only measure of comparison can be misleading \cite{Zitzler2003}, as it measures three (i.e. convergence, spread and cardinality) of the four aspects used to assess the quality of the calculated Pareto front \cite{Li2019}, it leaves out the fourth: uniformity. Therefore, we also consider the number of non dominated solutions found that lie in the reference (i.e. exact) Pareto front, as measure of coincidence in form. Although there are other indicators to measure the quality of the Pareto front ~\cite{ali2020} that quantify these aspects, the process of selecting quality indicators for the evaluation of non-dominated solutions sets is beyond the scope of this work. Thus, algorithms variance is checked by studying the sets of solutions found, in terms of hypervolume and number of solutions matching the reference (i.e. exact) Pareto front, after doing a fixed number of consecutive executions for the same problem instance.

\subsection{Methodology} \label{subsec:methodology}
The goal of the experiments is twofold: validate whether incorporating requirements interactions enhance the search process and evaluate the behaviour and performance of the EDAs. All algorithms were programmed in R and the NRP instances were solved using a 2,6 GHz CPU machine, with two kernels and 8 GB of RAM. The five stages protocol followed in the experiments consisted in:

\begin{enumerate}
    \item \emph{Construction of the reference front}. Consider a NRP instance, i.e. a data set and a fixed budget constraint. If its size (i.e. number of requirements) allows, apply exact search algorithms and find the exact Pareto front. Otherwise, divide the NRP instance ad hoc in several sub-problems, so each one contains approximately the same number of requirements (i.e. that can be handled by one of the exact algorithms) affected by dependencies that are not shared with other sub-problems %\color{magenta}
    (i.e. the sub-problems are disjoint in terms of requirements and interactions). Then apply exact search algorithms and find a set of %\color{magenta}
    valid solutions (in terms of interactions and effort limit) \color{black}
    for each NRP sub-problem. Afterwards, %\color{magenta}
    compute the Cartesian product of the valid solution sets \color{black} (i.e. partial valid solutions are concatenated together into a solution for the original problem, %\color{magenta}
    discarding those that exceed the effort limit\color{black}). Finally, %\color{magenta}
    from the set of valid solutions to the problem, \color{black} obtain the set of non-dominated solutions, i.e. the exact Pareto front.

    \item \emph{Capture  requirements interactions}. For each data set we construct the interaction graph for the instance following the process described in Section \ref{subsec:interactions_transformation}.
    
    \item \emph{Search algorithms set up}. Before running the EDA, some parameters need to be set up. We have to choose the number of iterations that will be performed, the population size, the number of consecutive iterations without changes between populations, the method that will be used to initialize the population and the sampling method to apply. In the experiments, the number of iterations is kept fixed and equal to 100 for small and medium size NRP instances, while is set to five times the number of requirements of the large size NRP instance, so we can get an insight on parameters influence. The population size is chosen to be 5 times the number of requirements in the instance, the number of consecutive iterations without changes in the population is selected as one tenth of the number of iterations. The population can be initialized randomly, using probabilistic logic sampling or maximum probability search. In all cases, the sampling method used is probabilistic logic sampling.
    
    \item \emph{Search algorithms execution}. Once the parameters have been chosen, the number of times the EDA is run is one quarter of the number of iterations set for the NRP instance.  In each execution, the front found, its hypervolume, the number of solutions it contains, the number of solutions it shares with the reference front, the number of iterations carried out and the time it has taken are stored for later evaluation. To get an idea of how an algorithm behaves, the 5 number summary statistics plus the mean of these values are reported and visualized.
    
    \item \emph{Results evaluation}. We use the hypervolume and the number of solutions found that lie in the reference Pareto front to check the performance of the algorithms. The stability of the results is measured using the coefficient of variation (i.e. standard deviation divided by the average) of hypervolume and number of coincident solutions. If it does not exceed the amount of 5\%, the results can be considered stable. 
    
    At a first level, the results obtained after several independent executions of the algorithms are analyzed \emph{within each instance of NRP}. Thus, to find out whether there are statistically significant differences in the hypervolume and the number of coincident solutions in a given NRP instance, we perform a Wilcoxon rank-sum test with Holm's correction (to compare the results of the runs of two algorithms) or a Kruskal-Wallis test (when comparing the results of the runs of three algorithms). When the Kruskal-Wallis test suggests that statistically significant differences exist, a post-hoc analysis using the Connover-Iman test is performed to detect which algorithms behave differently. 
    The version of the Wilcoxon test used is the one offered by the \emph{coin} R-package \cite{hothorn2008} %that allows to obtain p-values when ties occur
    , while for the Kruskal-Wallis and Conover-Iman tests the versions offered by the \emph{PMCMR-plus} R-package \cite{pohlert2021} have been used.
     
    At a second level, \emph{the performance of the algorithms is compared over all NRP instances} \cite{garcia2010, santafe2015}. As reliable and commensurable estimates of the algorithms' performance on each instance, we compute the average percentage of hypervolume and of the number of coincident solutions. Then, a Wilconxon signed-rank test is applied to see if there is evidence of statistically significant differences in the performance of two algorithms \cite{demsar2006, garcia2008}. Whereas, if three algorithms were implied in the comparison, we apply the Friedman aligned rank test, followed by a post-hoc analysis with the Friedman test using Finner's correction to check statistical differences by pairs \cite{garcia2010, santafe2015}. The version of the Wilcoxon signed-rank test used is the one offered by the \emph{exactRankTests} R-package \cite{hothorn2021}, while for the Friedman aligned rank and the post-hoc Friedman tests the \emph{scmamp} R-package \cite{borja2016} have been used.
    
\end{enumerate}

%%%%% NEW HYPERVOLUMEN SUMARY
\begin{table}[htpb]
\caption{Hypervolume of the exact Pareto fronts and hypervolume (mean, standard deviation and coefficient of variation) of the fronts found by the EDAs.}
\label{tab:hypervolume}
%\scalebox{0.96}{
\resizebox{\textwidth}{!}{
\begin{tabular}{p{5ex}rrrrlrrllll}
\hline
\multicolumn{2}{c}{\textit{\textbf{instance}}} &
  \multicolumn{1}{c}{\textit{\textbf{exact}}} &
  \multicolumn{2}{c}{\textit{\textbf{random}}} &
   &
  \multicolumn{2}{c}{\textit{\textbf{pls}}} &
   &
  \multicolumn{2}{c}{\textit{\textbf{maxprob}}} &
   \\ \cline{4-12} 
 &
  \multicolumn{1}{l}{} &
  \multicolumn{1}{l}{} &
  \multicolumn{1}{c}{\textbf{mean}} &
  \multicolumn{1}{c}{\textbf{sd}} &
  \multicolumn{1}{c}{\textbf{cv}} &
  \multicolumn{1}{c}{\textbf{mean}} &
  \multicolumn{1}{c}{\textbf{sd}} &
  \multicolumn{1}{c}{\textbf{cv}} &
  \multicolumn{1}{c}{\textbf{mean}} &
  \multicolumn{1}{c}{\textbf{sd}} &
  \multicolumn{1}{c}{\textbf{cv}} \\ \hline
       & 25  & 7905    & 7899.8    & 17.14    & 0.0022 & 7904.28  & 1.99     & 0.0003  & \multicolumn{1}{r}{7904.52}  & \multicolumn{1}{r}{1.66}  & 0.0002 \\
nrp20  & 43  & 18629   & 18293.6   & 145.55   & 0.0080 & 18424.48 & 139.04   & 0.0075 & \multicolumn{1}{r}{18603.72} & \multicolumn{1}{r}{34.25} & 0.0018 \\
       & 60  & 31165   & 30193.8   & 266.78    & 0.0088 & 30547.60 & 276.83   & 0.0091 & \multicolumn{1}{r}{30984.68} & \multicolumn{1}{r}{72.30} & 0.0023 \\ \hline
       & 107 & 177129  & 176999.0  & 59.90    & 0.0003 & 176349.9 & 445.31   & 0.0025 &                              &                           &        \\
nrp50  & 179 & 417245  & 413237.1  & 1872.00  & 0.0045 & 415977.6 & 586.46   & 0.0014 &                              &                           &        \\
       & 268 & 857937  & 849744.2  & 606.31   & 0.0007 & 849256.8 & 507.68   & 0.0006 &                              &                           &        \\ \hline
       & 311 & 249466  & 243347.7  & 679.33   & 0.0028 & 216512.5 & 3623.65  & 0.0167 &                              &                           &        \\
nrp100 & 519 & 583326  & 547308.1  & 3473.39  & 0.0063 & 491828.4 & 8734.29  & 0.0178 &                              &                           &        \\
       & 778 & 1243000 & 1007867.0 & 10507.80 & 0.0104 & 998067.7 & 12478.11 & 0.0125 &                              &                           &        \\ \hline
\end{tabular} }
\end{table}

%%%% NEW COINCIDENT SOLUTIONS SUMARY
\begin{table}[htpb]
\caption{Number of solutions on exact Pareto fronts and number of coincident solutions (mean, standard deviation and coefficient of variation) of the fronts found by the EDAs.}
\label{tab:coincident_solutions}
\resizebox{\textwidth}{!}{
\begin{tabular}{lrrrllrlllll}
\hline
\multicolumn{2}{c}{\textit{\textbf{instance}}} &
  \multicolumn{1}{c}{\textit{\textbf{exact}}} &
  \multicolumn{3}{c}{\textit{\textbf{random}}} &
  \multicolumn{3}{c}{\textit{\textbf{pls}}} &
  \multicolumn{3}{c}{\textit{\textbf{maxprob}}} \\ \cline{4-12} 
 &
  \multicolumn{1}{l}{} &
  \multicolumn{1}{l}{} &
  \multicolumn{1}{c}{\textit{\textbf{mean}}} &
  \multicolumn{1}{c}{\textit{\textbf{sd}}} &
  \multicolumn{1}{c}{\textit{\textbf{cv}}} &
  \multicolumn{1}{c}{\textit{\textbf{mean}}} &
  \multicolumn{1}{c}{\textit{\textbf{sd}}} &
  \multicolumn{1}{c}{\textit{\textbf{cv}}} &
  \multicolumn{1}{c}{\textit{\textbf{mean}}} &
  \multicolumn{1}{c}{\textit{\textbf{sd}}} &
  \multicolumn{1}{c}{\textit{\textbf{cv}}} \\ \hline
                & 25  & 19  & 18.32  & 0.69 & 0.0377 & 18.36  & 0.70 & 0.0381 & 18.92 & 0.28 & 0.0146 \\
\textit{nrp20}  & 43  & 27  & 23.08  & 0.70 & 0.0304 & 23.80  & 0.91 & 0.0384 & 25.56 & 0.87 & 0.0340 \\
                & 60  & 32  & 23.60  & 1.15 & 0.0489 & 24.76  & 1.67 & 0.0673 & 27.52 & 1.39 & 0.0504 \\ \hline
                & 107 & 69  & 66.28  & 1.40 & 0.0211 & 57.60  & 2.53 & 0.0440 &       &      &        \\
\textit{nrp50}  & 179 & 122 & 92.80  & 4.54 & 0.0489 & 98.44  & 4.41 & 0.0448 &       &      &        \\
                & 268 & 199 & 150.64 & 5.84 & 0.0388 & 142.64 & 4.72 & 0.0331 &       &      &        \\ \hline
                & 311 & 256 & 90.36  & 5.04 & 0.0558 & 136.91 & 3.71 & 0.0271 &       &      &        \\
\textit{nrp100} & 519 & 425 & 1.74   & 1.01 & 0.5776 & 193.31 & 3.76 & 0.0195 &       &      &        \\
                & 778 & 542 & 0.25   & 0.58 & 2.3274 & 74.79  & 7.87 & 0.1052 &       &      &        \\ \hline
\end{tabular} }
\end{table}

%%%% NEW EXECUTION TIMES SUMARY
\begin{table}[htpb]
\caption{Execution times (in seconds) of exact algorithms, EDAs (mean and standard deviation) and percentage of change when independencies are taken into account in the branch and bound algorithm.}
\label{tab:time}
\resizebox{\textwidth}{!}{
\begin{tabular}{lrrrrrrrrll}
\hline
\multicolumn{2}{c}{\textit{\textbf{instance}}} &
  \multicolumn{1}{c}{\textit{\textbf{brute}}} &
  \multicolumn{1}{c}{\textit{\textbf{branch \&}}} &
  \multicolumn{1}{l}{\textit{\textbf{change \%}}} &
  \multicolumn{2}{c}{\textit{\textbf{random}}} &
  \multicolumn{2}{c}{\textit{\textbf{pls}}} &
  \multicolumn{2}{c}{\textit{\textbf{maxprob}}} \\ \cline{6-11} 
 &
  \multicolumn{1}{l}{} &
  \multicolumn{1}{c}{\textit{\textbf{force}}} &
  \multicolumn{1}{c}{\textit{\textbf{bound}}} &
  \multicolumn{1}{l}{} &
  \multicolumn{1}{c}{\textit{\textbf{mean}}} &
  \multicolumn{1}{c}{\textit{\textbf{sd}}} &
  \multicolumn{1}{c}{\textit{\textbf{mean}}} &
  \multicolumn{1}{c}{\textit{\textbf{sd}}} &
  \multicolumn{1}{c}{\textit{\textbf{mean}}} &
  \multicolumn{1}{c}{\textit{\textbf{sd}}} \\ \hline
 &
  25 &
  13.83 &
  5.65 &
  -59.15 &
  5.33 &
  0.80 &
  5.32 &
  0.97 &
  \multicolumn{1}{r}{645.46} &
  \multicolumn{1}{r}{16.20} \\
\multicolumn{1}{c}{\textit{nrp20}} &
  43 &
  111.27 &
  83.36 &
  -25.08 &
  6.29 &
  1.35 &
  6.80 &
  1.39 &
  \multicolumn{1}{r}{8656.38} &
  \multicolumn{1}{r}{190.42} \\
 &
  60 &
  228.72 &
  192.30 &
  -15.92 &
  7.36 &
  1.62 &
  8.07 &
  1.80 &
  \multicolumn{1}{r}{9726.51} &
  \multicolumn{1}{r}{72.38} \\ \hline
 &
  107 &
  1065.76 &
  656.93 &
  -38.08 &
  78.23 &
  1.42 &
  79.50 &
  6.46 &
   &
   \\
\multicolumn{1}{c}{\textit{nrp50}} &
  179 &
  1410.37 &
  813.31 &
  -42.33 &
  74.49 &
  1.05 &
  78.34 &
  1.32 &
   &
   \\
 &
  268 &
  1463.12 &
  866.19 &
  -40.80 &
  75.48 &
  0.87 &
  77.67 &
  1.00 &
   &
   \\ \hline
 &
  311 &
  \multicolumn{1}{l}{} &
  \multicolumn{1}{l}{} &
  \multicolumn{1}{l}{} &
  2503.51 &
  20.46 &
  2724.45 &
  44.16 &
   &
   \\
\multicolumn{1}{c}{\textit{nrp100}} &
  519 &
  \multicolumn{1}{l}{} &
  \multicolumn{1}{l}{} &
  \multicolumn{1}{l}{} &
  2378.69 &
  12.52 &
  2678.42 &
  14.48 &
   &
   \\
 &
  778 &
  \multicolumn{1}{l}{} &
  \multicolumn{1}{l}{} &
  \multicolumn{1}{l}{} &
  2431.41 &
  12.24 &
  2494.10 &
  11.65 &
   &
   \\ \hline
\end{tabular} }
\end{table}

\vspace{24pt}
\subsection{Results and discussion}
In this section, following the proposed methodology, the results (i.e. hypervolume and number of coincident solutions) of the independent runs of the algorithms in the different NRP instances are collected and analysed to study their behaviour and to see if there are differences between them. To do so, we first focus on the analysis of the results obtained in the instances associated to each NRP problem (i.e. NRP20, NRP50 and NRP100).  Then, we analyse and compare the overall performance of the algorithms on all the NRP instances in order to identify if any of them could be better than the others.

% NRP20 Hypervolume and coincident solutions distributions
\begin{figure} [htbp]
 \centering
      \begin{subfigure}[b]{0.48\textwidth}
    %    \centering
    \includegraphics[width=\textwidth]{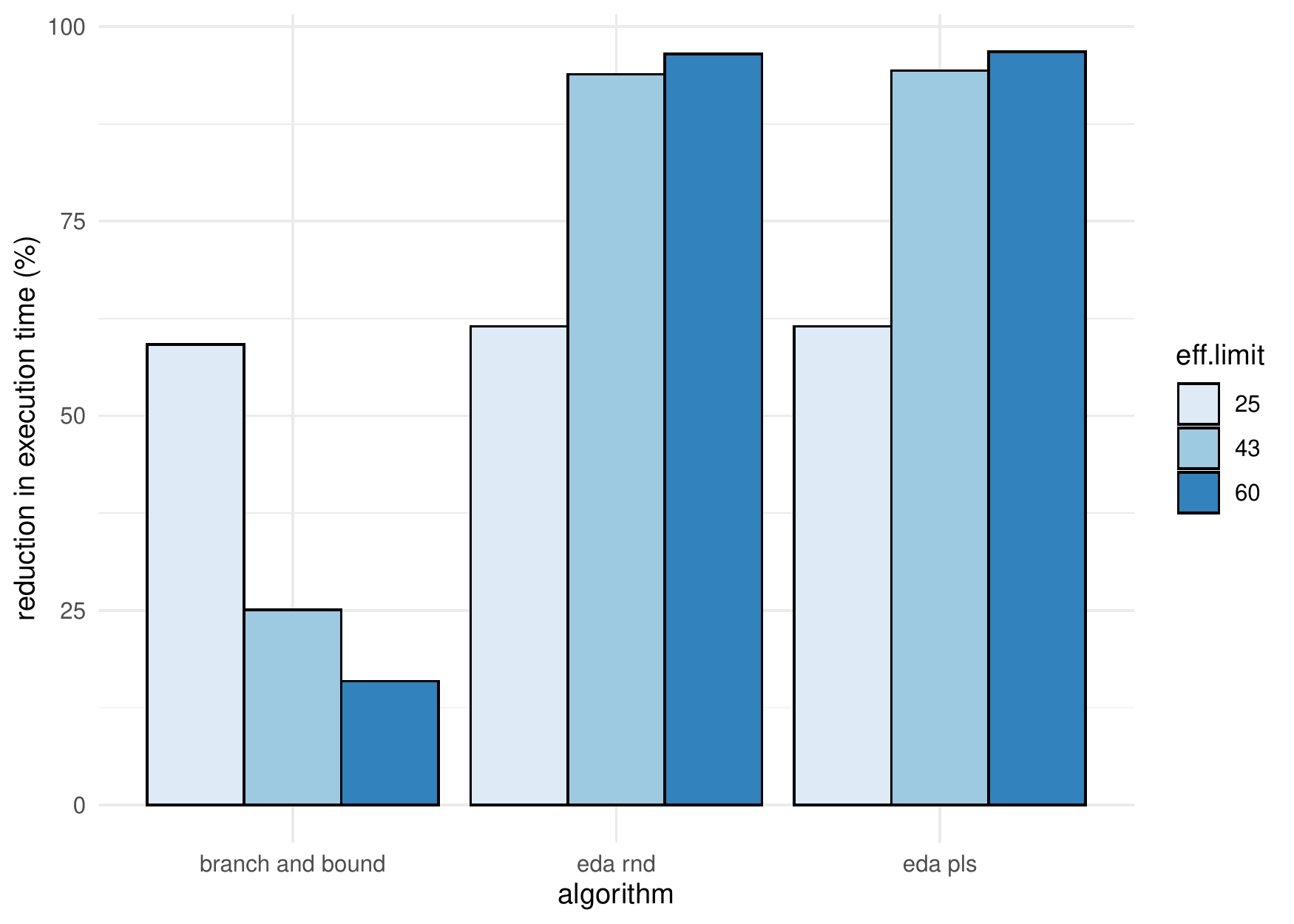}
      \caption{%Percentage of r
      Exec. times reduction with respect to brute force search.}
      \label{fig:nrp20_time_reduction}
     \end{subfigure}
     \begin{subfigure}[b]{0.46\textwidth}
      %   \centering
       \includegraphics[width=\textwidth]{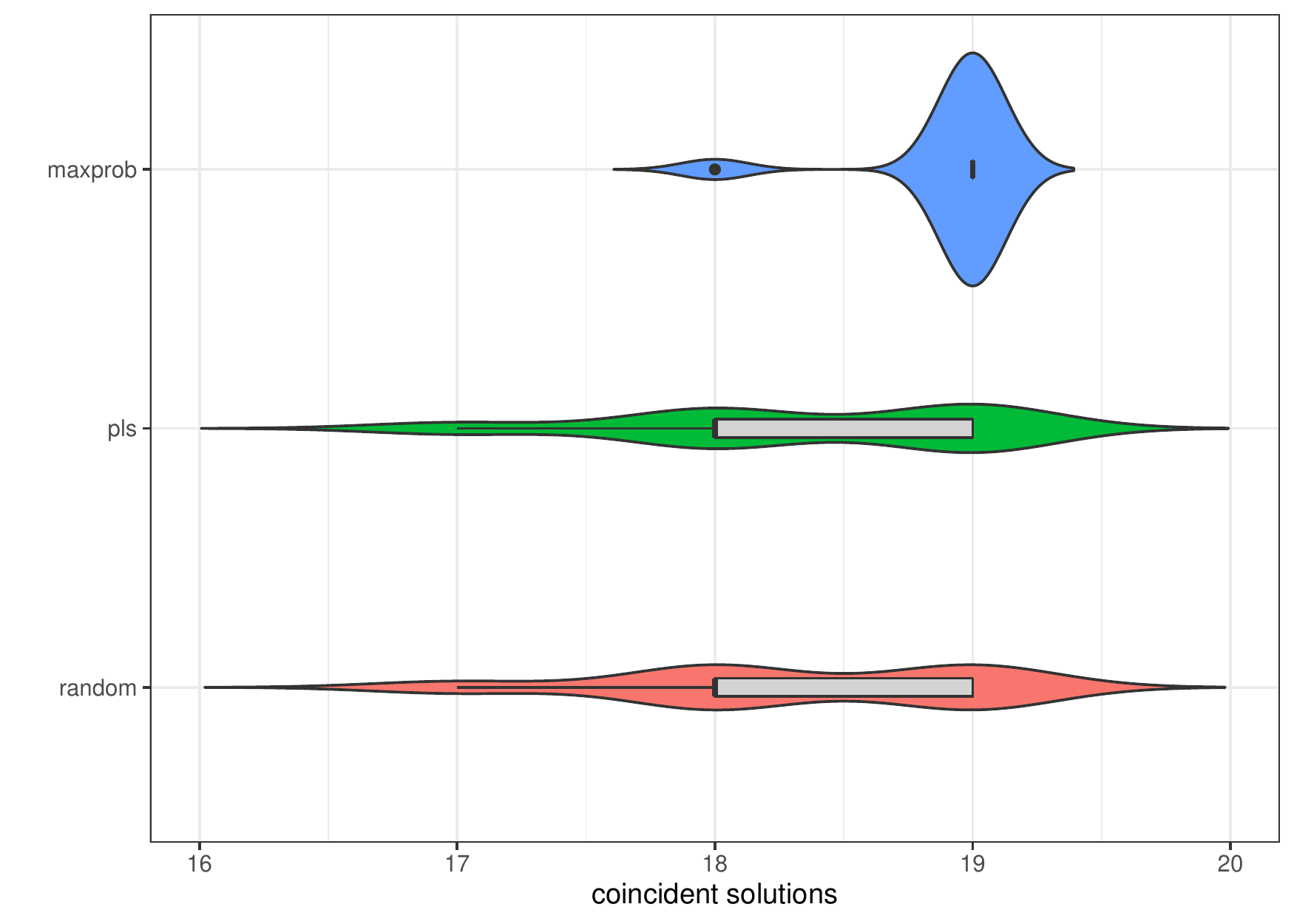}
     \caption{Number of coincident solutions effort 25}
       \label{fig:nrp20_25_coincident_sols_distribution}
        \end{subfigure}
    \begin{subfigure}[b]{0.46\textwidth}
   %      \centering
         \includegraphics[width=\textwidth]{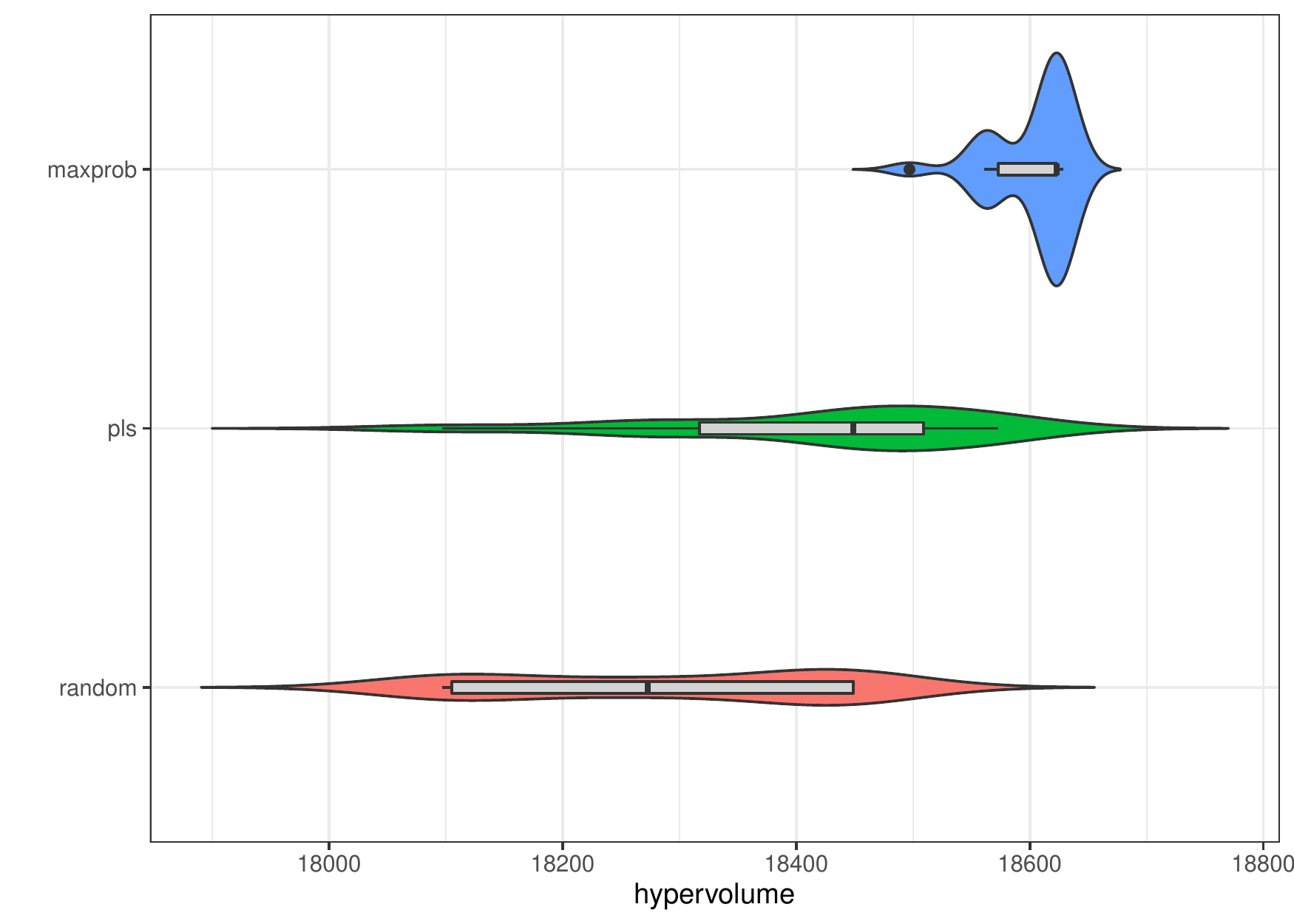}
         \caption{  Hypervolume effort 43}
         \label{fig:nrp20_43_hypervol}
     \end{subfigure}
     \hfill
     \begin{subfigure}[b]{0.46\textwidth}
   %      \centering
         \includegraphics[width=\textwidth]{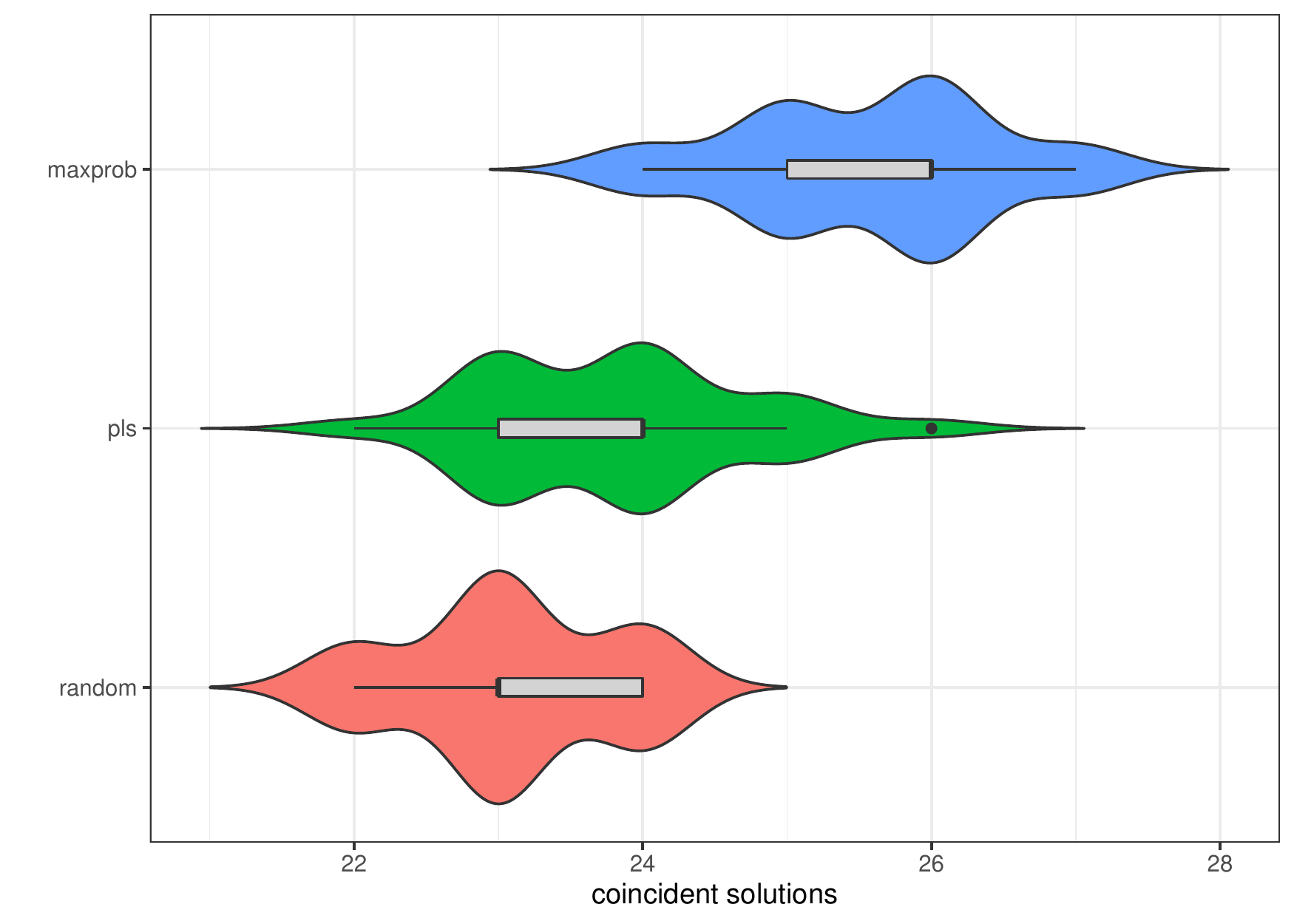}
         \caption{Number of coincident solutions effort 43}
         \label{fig:nrp20_43_coincident_sols}
     \end{subfigure}
      \begin{subfigure}[b]{0.46\textwidth}
 %        \centering
         \includegraphics[width=\textwidth]{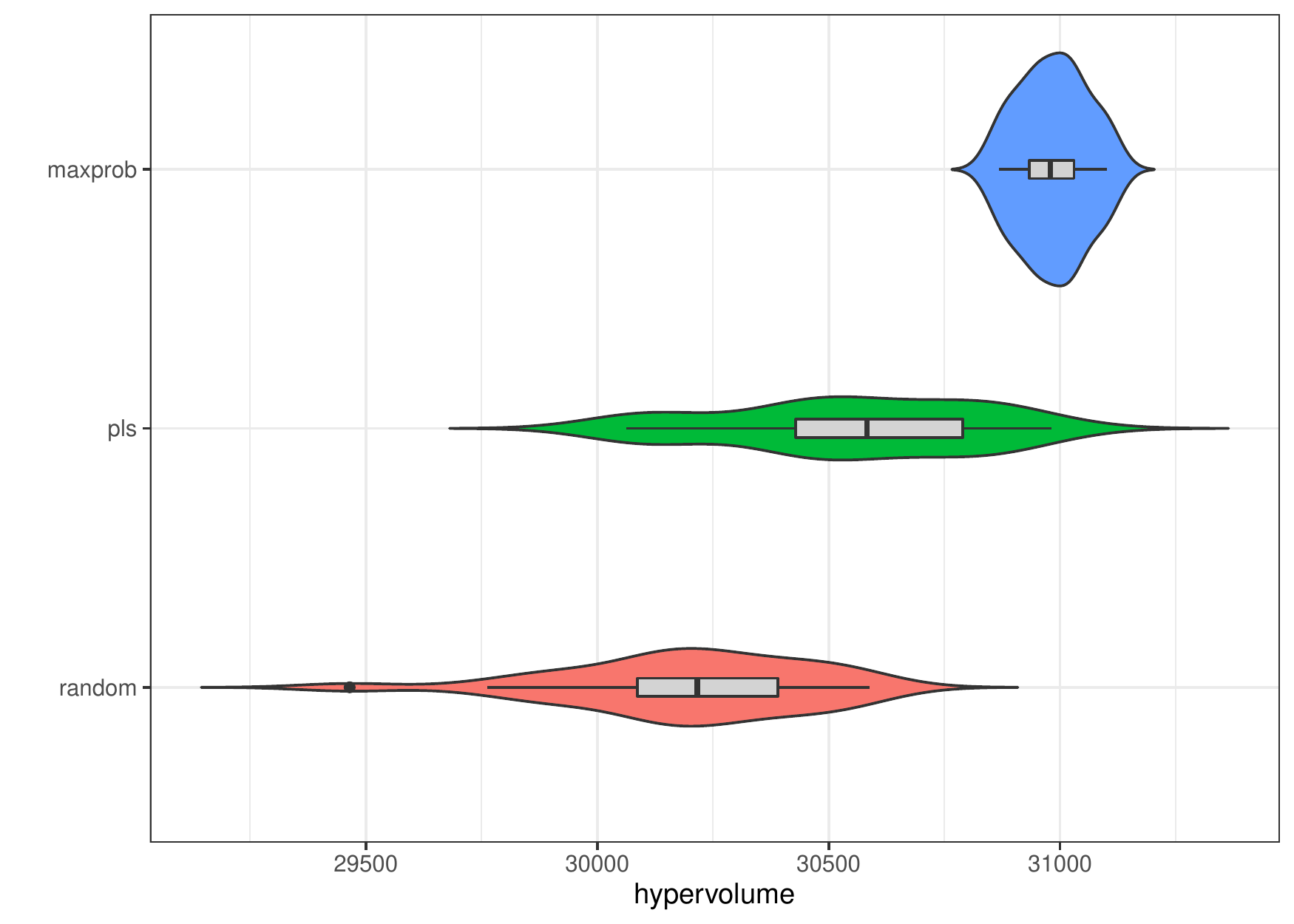}
         \caption {Hypervolume effort 60}
         \label{fig:nrp20_60_hypervol}
     \end{subfigure}
     \hfill
     \begin{subfigure}[b]{0.46\textwidth}
  %       \centering
         \includegraphics[width=\textwidth]{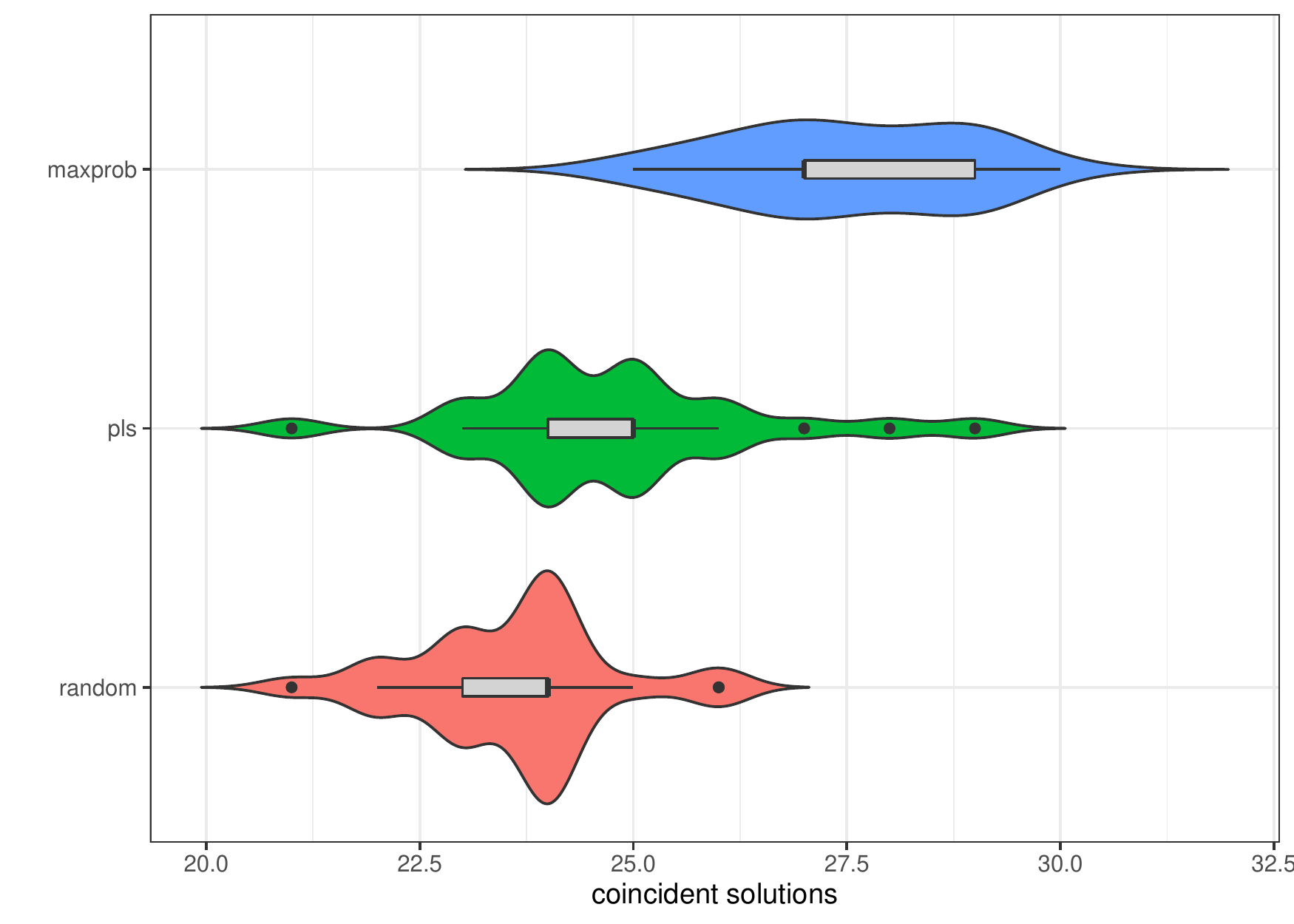}
         \caption{Number of coincident solutions effort 60}
         \label{fig:nrp20_60_coincident_sols}
     \end{subfigure}
        \caption{NRP20 Comparison of the EDA's distributions of hypervolume and number of solutions found that lie in the reference front}
        \label{fig:nrp20_60_distributions}
\end{figure}
\color{black}

\subsubsection{NRP with 20 requirements}
Three instances of this problem were created by setting the effort limit to  25, 43 and 60. Tables \ref{tab:hypervolume} and \ref{tab:coincident_solutions} (see the \textit{exact} column) show, respectively for all instances, the hypervolume and the number of solutions of the exact Pareto fronts obtained considering interactions (branch-and-bound algorithm) or not (brute-force search algorithm).  The time used in their calculation is shown in Table \ref{tab:time} (see columns \textit{brute force} and \textit{branch \& bound}). The percentages of relative change (see column \textit{change \%} in Table \ref{tab:time}) indicate there is a reduction in execution time (\textbf{RQ2}), with respect to that of the brute force algorithm, when requirements interactions are included in the search process (\textbf{RQ1}). 

Even when population initialization is done randomly (\emph{rnd}) or using probabilistic logical sampling (\emph{pls}), the mean execution times of the algorithms also show a reduction (see Figure \ref{fig:nrp20_time_reduction}). Although, it could arise from the limitation on the number of executions to be performed, this is not the case. In most cases, the limit of 100 iterations set per run is not reached and, if it is reached (as when the effort limit is 60), it happens in less than 7 executions ($\leq 25$\%). However, when \emph{maxprob} initialization is applied in all the instances  (i.e., effort limit 25, 43, and 60), the mean execution times of the EDA are increased by a factor of 46.67, 77.79 and 42.53, respectively. This increase is due to how \emph{maxprob} builds the solutions. It generates the search tree by following the branch with the highest probability, at each instant. In this way, instead of moving along a branch until a solution is complete (i.e. all requirements have been considered to be included or not), \emph{maxprob} jumps from one branch to another following the maximum probability value and resembles a breath-first search  making it infeasible and impractical. Therefore, except for \emph{maxprob} initialization, even in approximate algorithms, as our EDA is, the answer to \textbf{RQ2} is positive: there is an improvement in the execution times when interactions are taken into account.

%results stability
As we do 25 EDA executions for each instance and each population initialization method,  the possible variations in the results (see mean and standard deviation values showed in columns \textit{mean} and \textit{sd} in Tables \ref{tab:hypervolume} and \ref{tab:coincident_solutions}) should be checked by calculating the coefficient of variation (CV) of the hypervolume and the number of solutions found that are on the Pareto front (see columns \textit{cv} in Tables \ref{tab:hypervolume}, \ref{tab:coincident_solutions}).  With regard to hypervolume results, the values of the coefficient of variation do not exceed an amount of 5\%, so the results can be considered stable. However, for the number of coincident solutions, the results are stable only in two instances \textit{nrp20} 25 and \textit{nrp20} 43. Next, let's study what we get in each particular instance with respect to hypervolume and number of coincident solutions:

%differences: standard α = 0.05 cutoff, the null hypothesis is rejected when p < .05 and not rejected when p > .05. 
\begin{itemize}
\item{\emph{NRP20, 25 effort limit.}}
For hypervolume distributions, the Kruskal-Wallis test (see Table \ref{tab:kruskal_wallis_NRP20}) does not detect significant differences ($p = 0.848 > 0.05$), whereas it does ($p = 6.62e-04 < 0.05$) for the distributions of coincident solutions (see Fig. \ref{fig:nrp20_25_coincident_sols_distribution}). The post hoc pairwise Conover-Iman test (see Table \ref{tab:post_hoc_pairwise_conover_NRP20}) indicates that there were significant differences between \emph{maxprob} and the other two initialization methods: \emph{random} ($p = 0.00085 < 0.05$) and probabilistic logic sampling ($p = 0.00216 < 0.05$).

\item{\emph{NRP20, 43 effort limit.}}
A Kruskal-Wallis test (see Table \ref{tab:kruskal_wallis_NRP20}) was carried out to compare the distributions of hypervolume (see Fig. \ref{fig:nrp20_43_hypervol}) and of number of coincident solutions (see Fig. \ref{fig:nrp20_43_coincident_sols}) of the Pareto fronts returned by the executions of the EDA with the population initialization methods used. There were found significant differences in hypervolume ($p = 8.35e-12 < 0.05$) and coincident solutions ($p = 1.31e-10 < 0.05$). Pairwise Conover-Iman post hoc tests were carried out to find out where the differences in hypervolume and coincident solutions were (see Table \ref{tab:post_hoc_pairwise_conover_NRP20}), indicating there were significant differences between all pairs (all \textit{p}-values are less than 0.05).

\item{\emph{NRP20, 60 effort limit.}}

In this case, the situation coincides with that found in the previous instance. There were significant differences (see Table \ref{tab:kruskal_wallis_NRP20}) in the distributions of hypervolume (see Fig. \ref{fig:nrp20_60_hypervol}), with $p = 2.868e-12 < 0.05$, and of number of coincident solutions (see Fig. \ref{fig:nrp20_60_coincident_sols}), with $p = 2.15e-10 < 0.05$. Pairwise Conover-Iman post hoc tests, were carried out to find out where the differences were (see Table \ref{tab:post_hoc_pairwise_conover_NRP20}) and indicate that there were significant differences in hypervolume and coincident solutions between all pairs (all p-values are less than 0.05).

\end{itemize}

%%% TESTS

\begin{table}[htpb]
\centering
\caption{Kruskal-Wallis tests \emph{p}-values for instances of NRP20.}
\label{tab:kruskal_wallis_NRP20}
\begin{tabular}{llrr}
\hline
               &             & \multicolumn{2}{c}{\textbf{Kruskal-Wallis tests}} \\ \hline
\multicolumn{2}{l}{\textbf{Instance}} & \textbf{Hypervolume}    & \textbf{Coincident solutions}    \\ \hline
               & 25          & 0.8476          & 6.20e-04                 \\
nrp20          & 43          & 8.35e-12       & 1.31e-10                \\
               & 60          & 2.86e-12       & 2.15e-10                \\ \hline
\end{tabular}
\end{table}

\begin{table}[htpb]
\centering
\caption{Post-hoc pairwise Conover-Iman tests p-values for NRP20 instances}
\label{tab:post_hoc_pairwise_conover_NRP20}
\begin{tabular}{lllllllll}
\hline
      &    &  & \multicolumn{6}{c}{\textbf{Pairwise Conover-Iman tests  (post-hoc)}} \\ \hline
\multicolumn{2}{c}{\textbf{Instance} }&  &  & \multicolumn{2}{c}{\textbf{Hypervolume} }&  & \multicolumn{2}{c}{\textbf{Coincident sols.}} \\ %\hline
      &    &  &         & random    & pls       &  & random    & pls       \\ \cline{1-2} \cline{4-6} \cline{8-9} 
      & 25 &  & pls     &           &           &  & 0.95566    &           \\
      &    &  & maxprob &           &           &  & 0.00085   & 0.00216    \\ \cline{4-6} \cline{8-9} 
nrp20 & 43 &  & pls     & 7.7e-05  &            &  & 0.0037    &           \\
      &    &  & maxprob & $<$ 2e-16  & $<$ 2e-16  &  & $<$ 2e-16 & 1.7e-09 \\ \cline{4-6} \cline{8-9} 
      & 60 &  & pls     & 1.4e-05  &           &  & 0.0025  &           \\
      &    &  & maxprob & $<$ 2e-16 & $<$ 2e-16  &  & $<$ 2e-16 & 8.4e-09 \\ \hline
\end{tabular}
\end{table}

In the last two instances, as the size of the search space increases due to the relaxation in the effort bound, EDA with \emph{pls} initialization returned best results in average than when \emph{random} initialization is used. Thus, the percentage differences in hypervolume were 0.71\% and 30.58\%, and in the number of solutions found that lie in the exact Pareto front were 3.12\% and 4.92\%, respectively.

% NRP50 Hypervolume and coincident solutions distributions
\begin{figure}[htbp]
     \centering
     \begin{subfigure}[b]{0.49\textwidth}
         \centering
         \includegraphics[width=\textwidth]{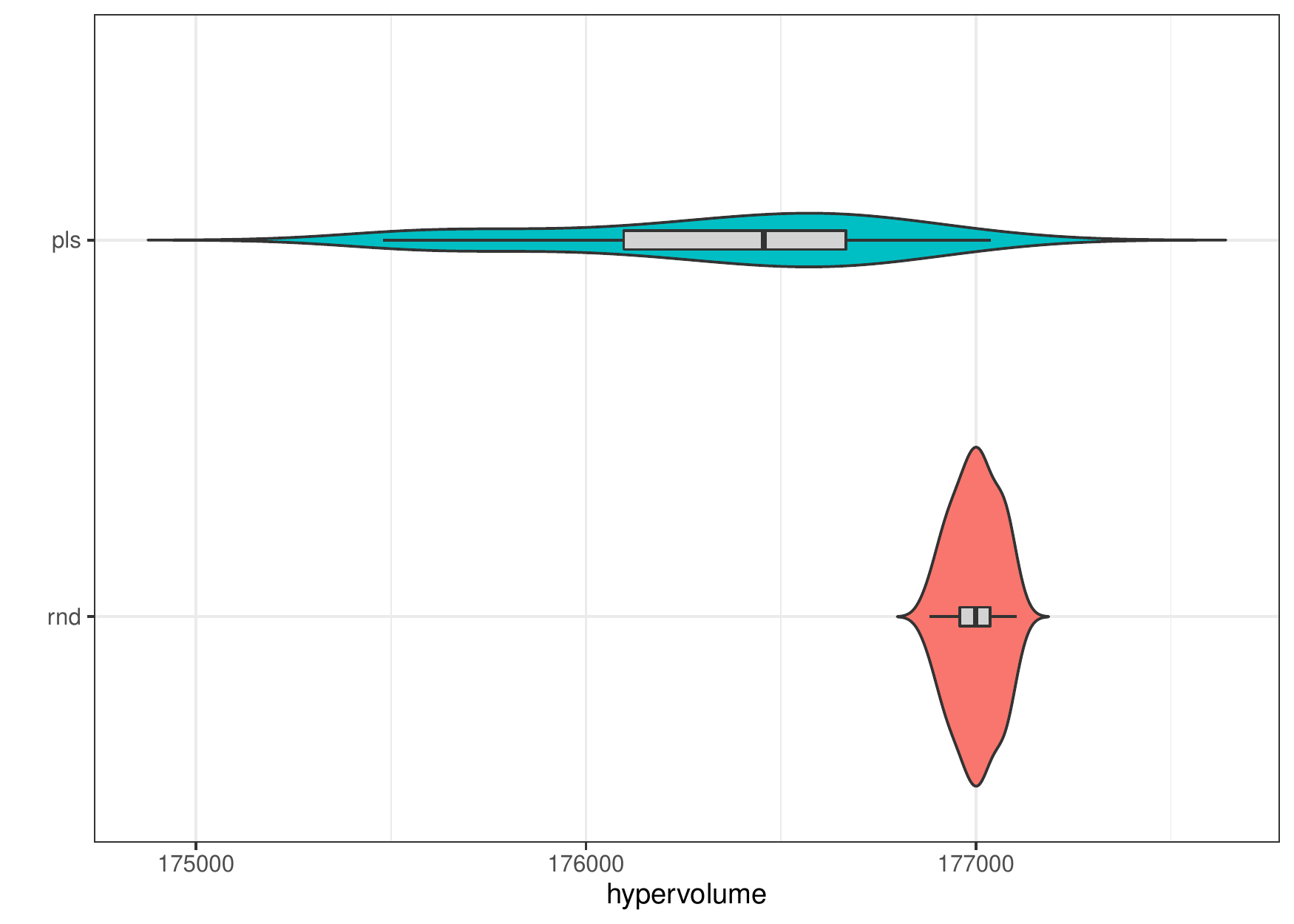}
         \caption{Hypervolume  107 effort}
         \label{fig:nrp50_107_hypervol}
     \end{subfigure}
     \hfill
     \begin{subfigure}[b]{0.49\textwidth}
         \centering
         \includegraphics[width=\textwidth]{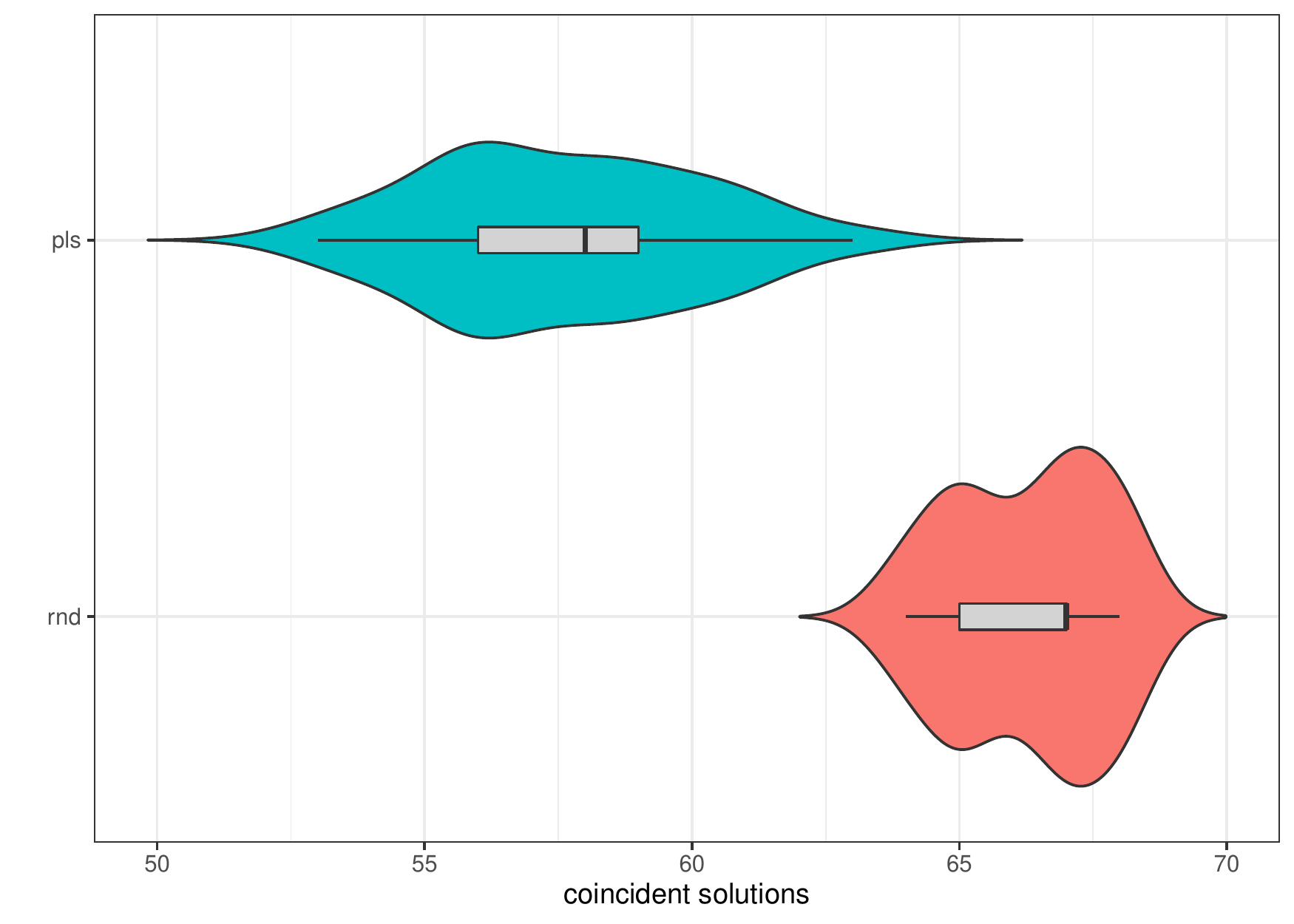}
         \caption{Number of coincident solutions  107 effort}
         \label{fig:nrp50_107_coincident_sols}
     \end{subfigure}

    \begin{subfigure}[b]{0.49\textwidth}
         \centering
         \includegraphics[width=\textwidth]{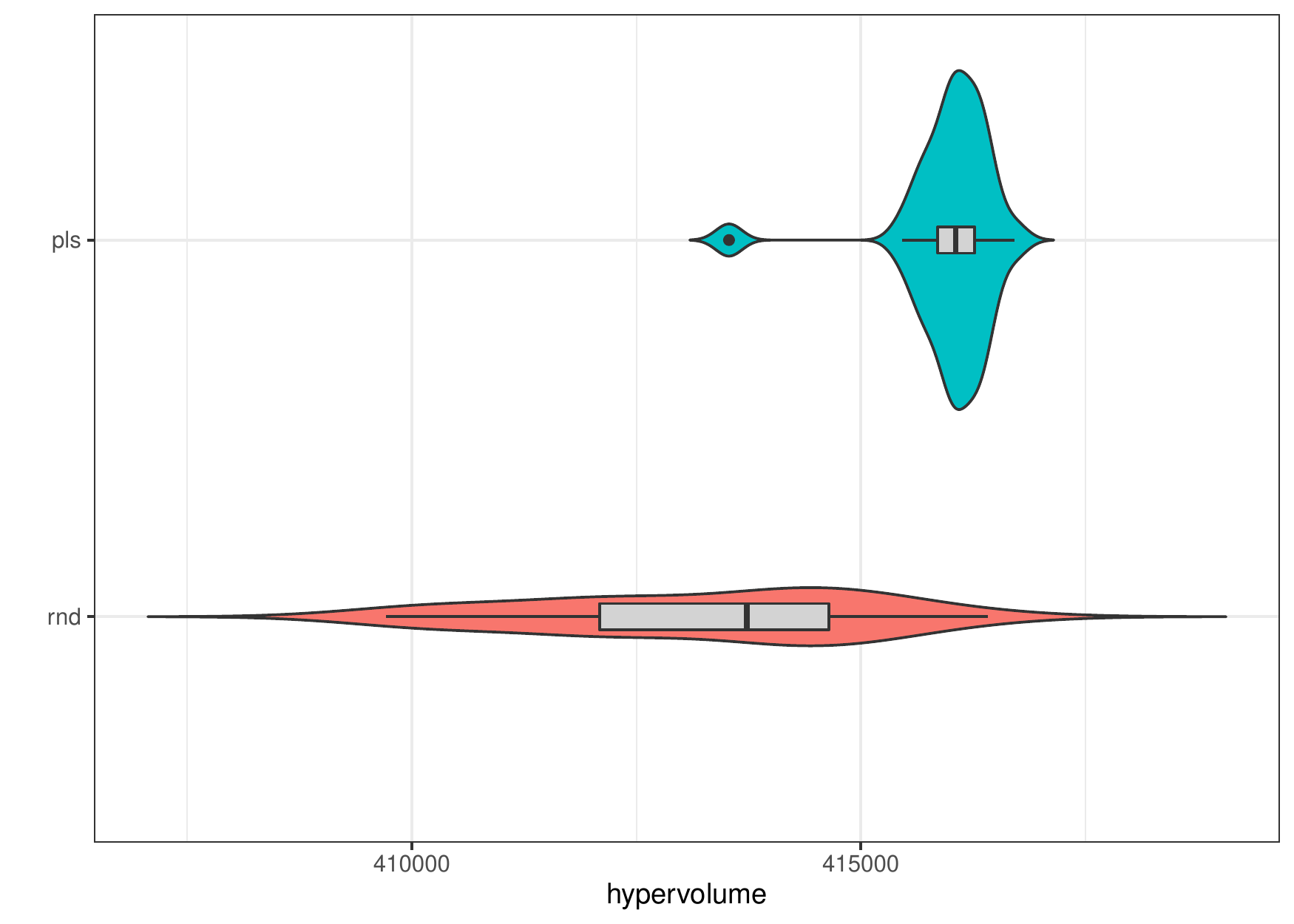}
         \caption {Hypervolume 179 effort}
         \label{fig:nrp50_179_hypervol}
     \end{subfigure}
     \hfill
     \begin{subfigure}[b]{0.49\textwidth}
         \centering
         \includegraphics[width=\textwidth]{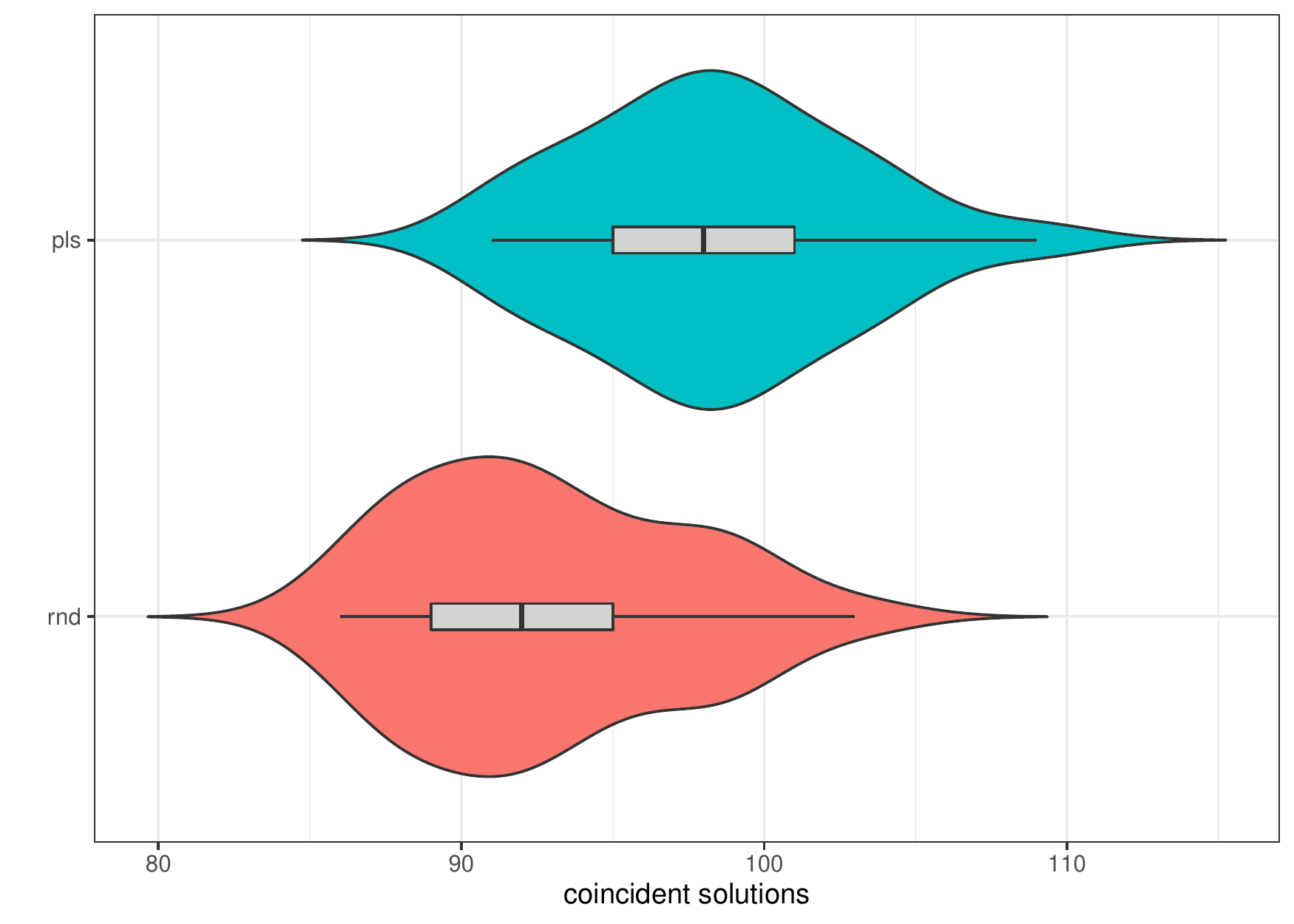}
         \caption{Number of coincident solutions 179 effort }
         \label{fig:nrp50_179_coincident_sols}
     \end{subfigure}
    \begin{subfigure}[b]{0.49\textwidth}
         \centering
         \includegraphics[width=\textwidth]{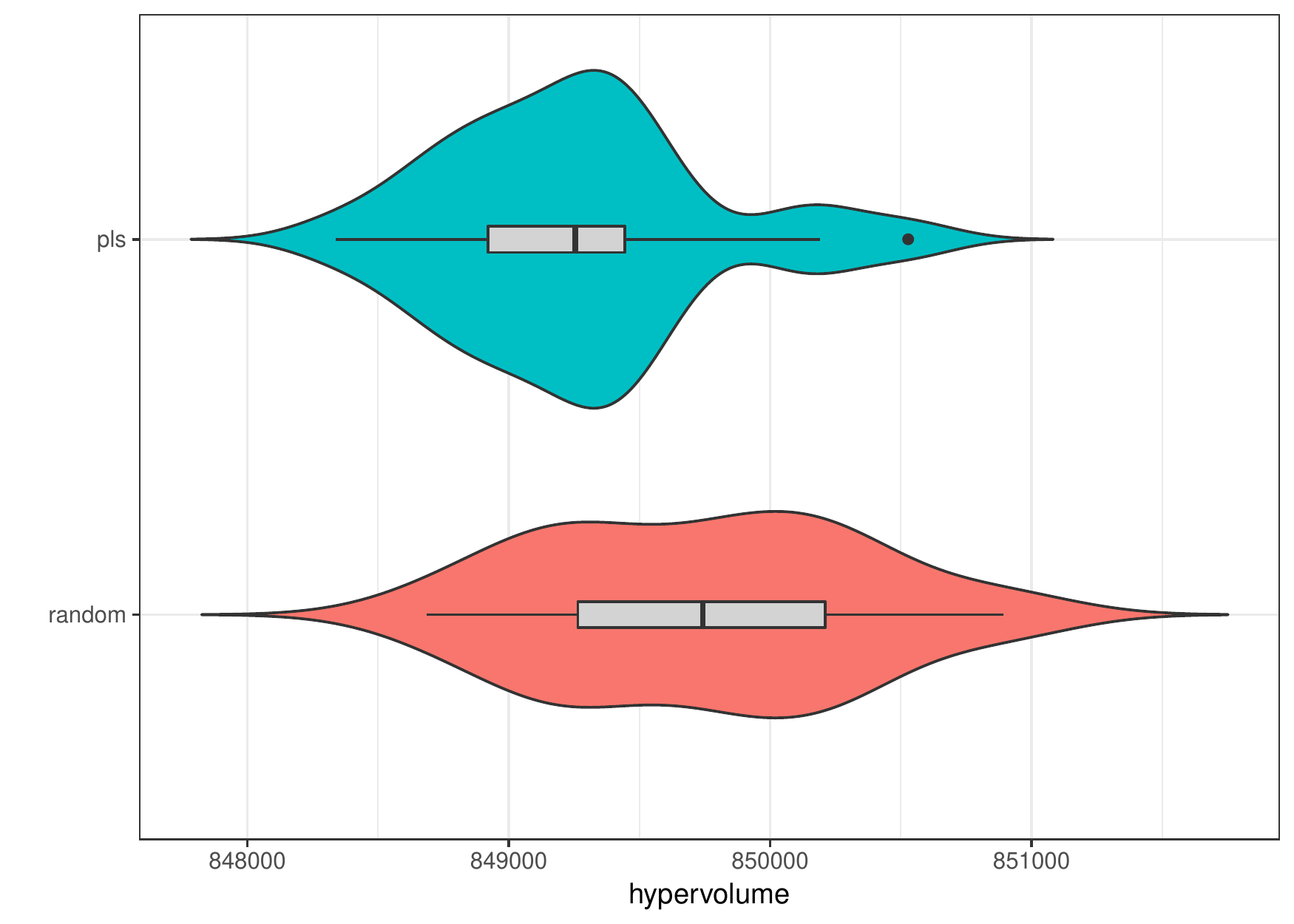}
         \caption{Hypervolume 268 effort}
         \label{fig:nrp50_268_hypervol}
     \end{subfigure}
     \hfill
     \begin{subfigure}[b]{0.49\textwidth}
         \centering
         \includegraphics[width=\textwidth]{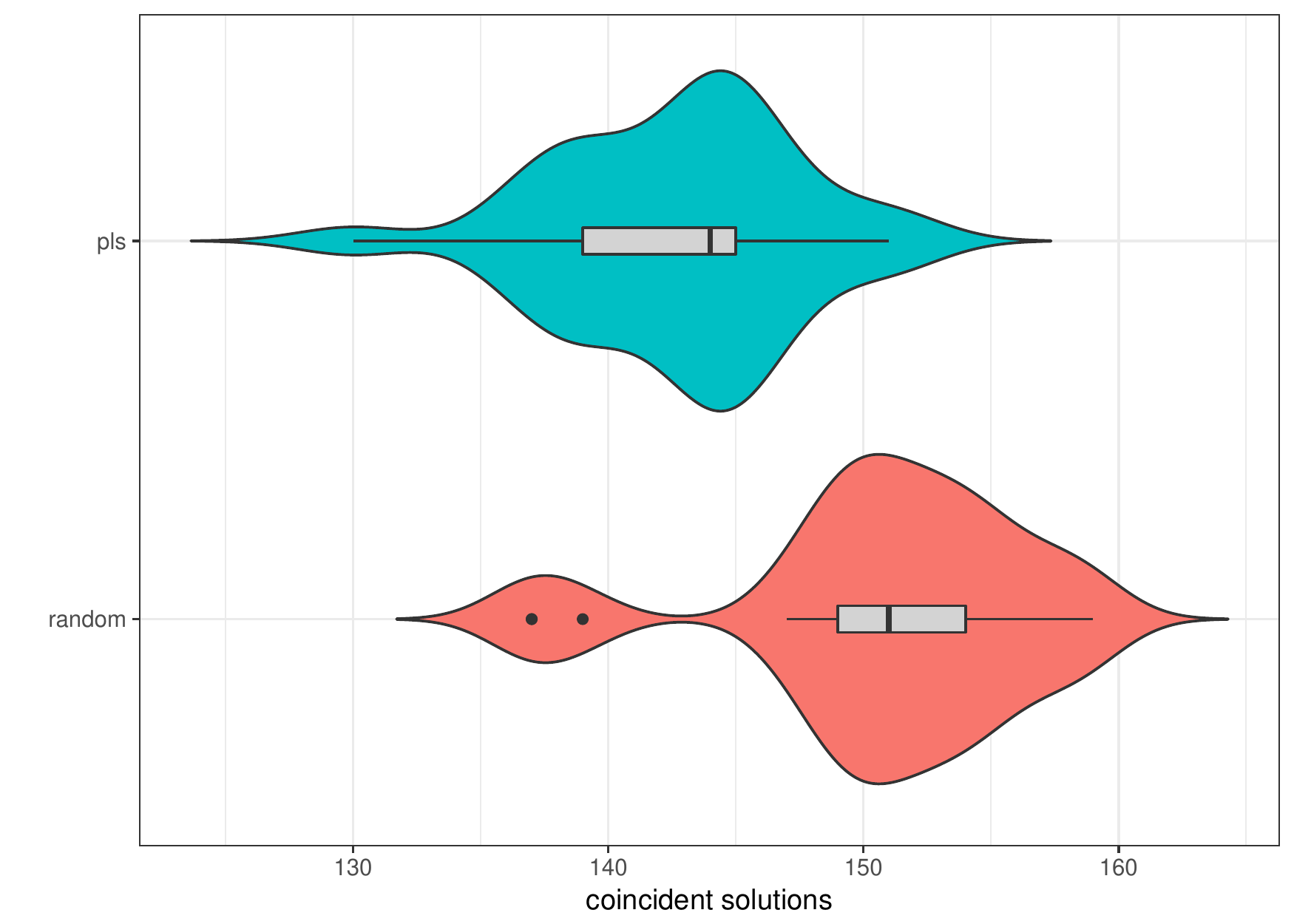}
         \caption{Number of coincident solutions 268 effort}
         \label{fig:nrp50_268_coincident_sols}
     \end{subfigure}    
          \caption{NRP50  comparison of the EDA's distributions of hypervolume and number of solutions found that lie in the reference front}   
        \label{fig:nrp50_107_distributions}   
\end{figure}

\subsubsection{NRP with 50 requirements}
As in the previous problem, we use three NRP instances corresponding to effort limits of 107, 179 and 268 (a 30\%,  50\% and 75\% of the total problem effort, respectively). Tables \ref{tab:hypervolume} and \ref{tab:coincident_solutions} (see the \emph{exact} column) show, respectively for all instances, the hypervolume and the number of solutions of the exact Pareto fronts obtained dividing ad hoc each instance into two sub-problems (see step 1 in subsection \ref{subsec:methodology}), applying the brute force search algorithm (without considering interactions) or the proposed branch and bound algorithm (interactions are taken into account) and combining the sets of solutions for each sub-problem. The time involved in their calculation is shown in Table \ref{tab:time} (see columns \textit{brute force} and \textit{branch \& bound}). The percentages of relative change (see column \textit{change \%} in Table \ref{tab:time}) indicate there is a reduction in execution time, with respect to that of the brute force algorithm, when requirements interactions are included in the search process (\textbf{RQ2}).

\emph{Maxprob} initialization is discarded due to its massive execution times, we only analyzed the results obtained by the EDA when \emph{random} and \emph{pls} initialization are applied. Figure \ref{fig:nrp50_107_distributions} shows the distributions of hypervolume (see mean and standard deviation values showed in columns \emph{mean} and \emph{sd} in Table \ref{tab:hypervolume}) and number of solutions found that are included in the reference front (see mean and standard deviation values showed in columns \emph{mean} and \emph{sd} in Table \ref{tab:coincident_solutions}). In all instances, the values of the coefficient of variation for hypervolume and number of coincident solutions (see columns \textit{cv} in Tables \ref{tab:hypervolume}, \ref{tab:coincident_solutions}) are less than 5\%, then EDA results can be considered stable. However, the Wilcoxon test detects significant differences ($p < 0.05$) between these distributions (see Table \ref{tab:wilcoxon_NRP50_NRP100}) in all instances.

\begin{table}[htpb]
\centering
\caption{Wilconxon tests p-values for instances of NRP50 and NRP100}
\label{tab:wilcoxon_NRP50_NRP100}
\begin{tabular}{llrr}
\hline
               &             & \multicolumn{2}{l}{\textbf{Wilcoxon tests (random-pls)}} \\ \hline
\multicolumn{2}{l}{\textbf{Instance} }& \textbf{Hypervolume}             & \textbf{Coincident sols.}       \\ \hline
               & 107         & 1.537e-08               & 1.1e-09                \\
nrp50          & 179         & 8.1e-08                 & 1.588e-04               \\
               & 268         & 7.633e-03                & 8.891e-06              \\ \cline{3-4} 
               & 311         & \textless 2.2e-16       & \textless 2.2e-16      \\
nrp100         & 519         & \textless 2.2e-16       & \textless 2.2e-16      \\
               & 778         & 5.722e-11               & \textless 2.2e-16      \\ \hline
\end{tabular}
\end{table}

The best results, in average, in the first and third instances were obtained using \emph{random} initialization. The percentage differences in hypervolume were 0.37\% and 0.06\%, and in the number of solutions found that lie in the exact Pareto front were 13.1\% and 5.31\%, respectively. While, in the second instance the best results were obtained using \emph{pls} initialization with differences of 0.66\% in hypervolume and 6.21\% in number of coincident solutions. Note the reduced execution time of the EDA algorithms, which was less than 90 seconds. Even with this time limitation, achieved by setting the number of iterations at 100, the approximate fronts found by the EDAs had a high quality when being compared with the reference fronts. This leads us to the answer to \textbf{RQ3}, EDAs are able to find good approximations to the Pareto fronts in a limited time.

% NRP100 Hypervolume and coincident solutions distributions
\begin{figure}[htbp]
     \centering
     \begin{subfigure}[b]{0.47\textwidth}
         \centering
         \includegraphics[width=\textwidth]{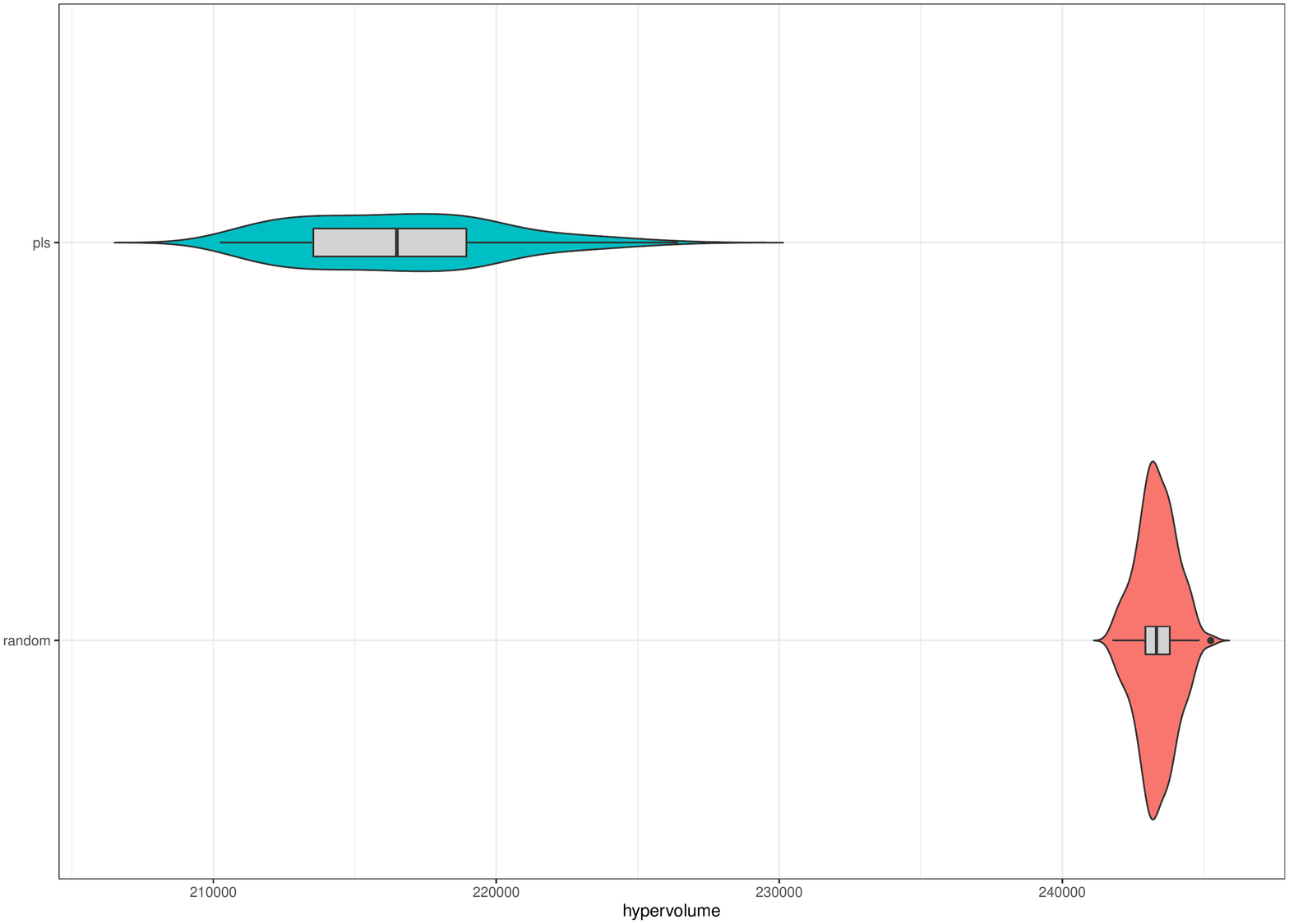}
         \caption{Hypervolume 311 effort}
         \label{fig:nrp100_311_hypervol}
     \end{subfigure}
     \hfill
     \begin{subfigure}[b]{0.47\textwidth}
         \centering
         \includegraphics[width=\textwidth]{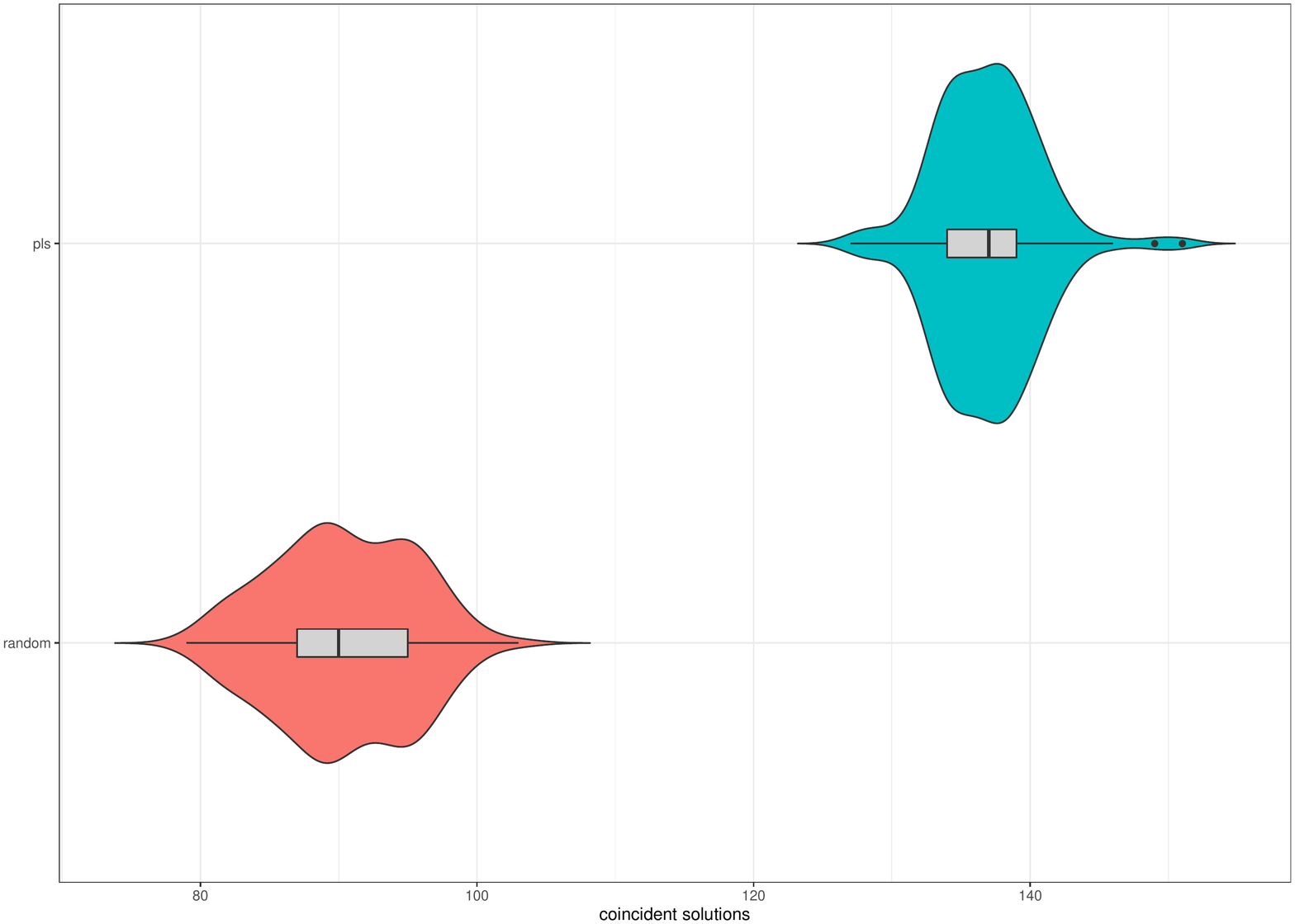}
         \caption{Number of coincident solutions 311 effort}
         \label{fig:nrp100_311_coincident_sols}
     \end{subfigure}
     \begin{subfigure}[b]{0.47\textwidth}
         \centering
         \includegraphics[width=\textwidth]{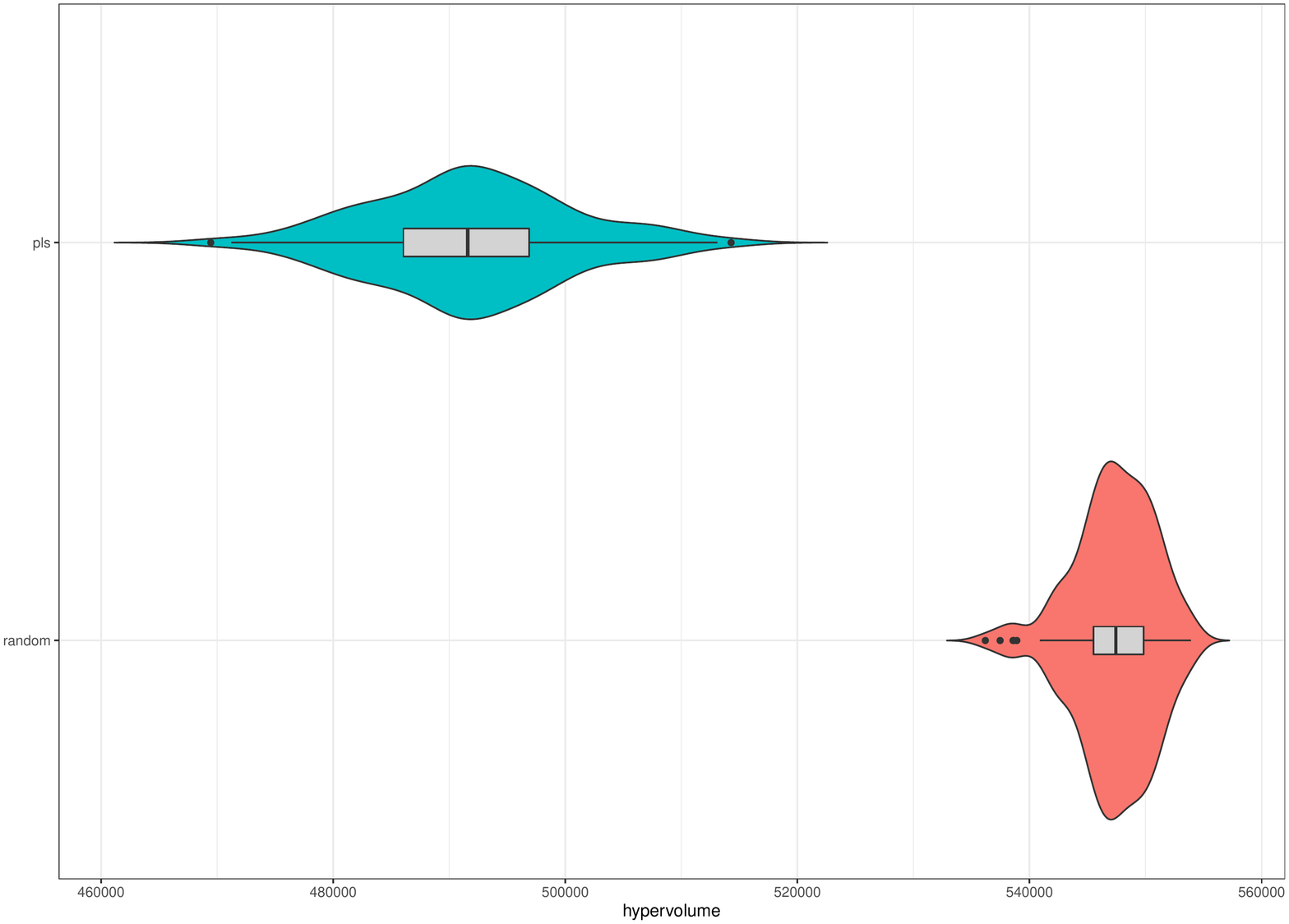}
         \caption{Hypervolume 519 effort}
         \label{fig:nrp100_519_hypervol}
     \end{subfigure}
     \hfill
     \begin{subfigure}[b]{0.47\textwidth}
         \centering
         \includegraphics[width=\textwidth]{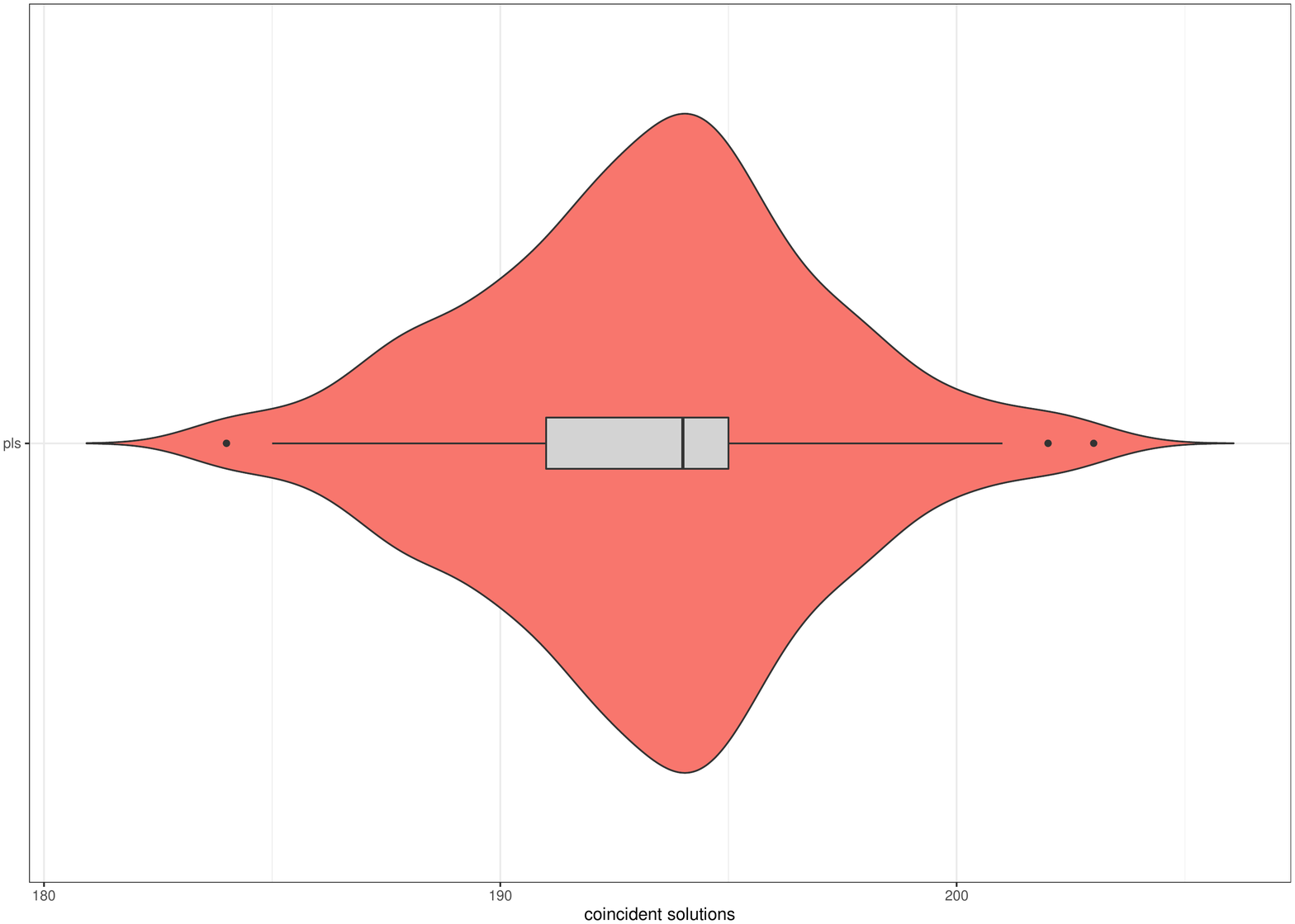}
         \caption{Number of coincident solutions 519 effort}
         \label{fig:nrp100_519_coincident_sols}
     \end{subfigure}
     \begin{subfigure}[b]{0.47\textwidth}
         \centering
         \includegraphics[width=\textwidth]{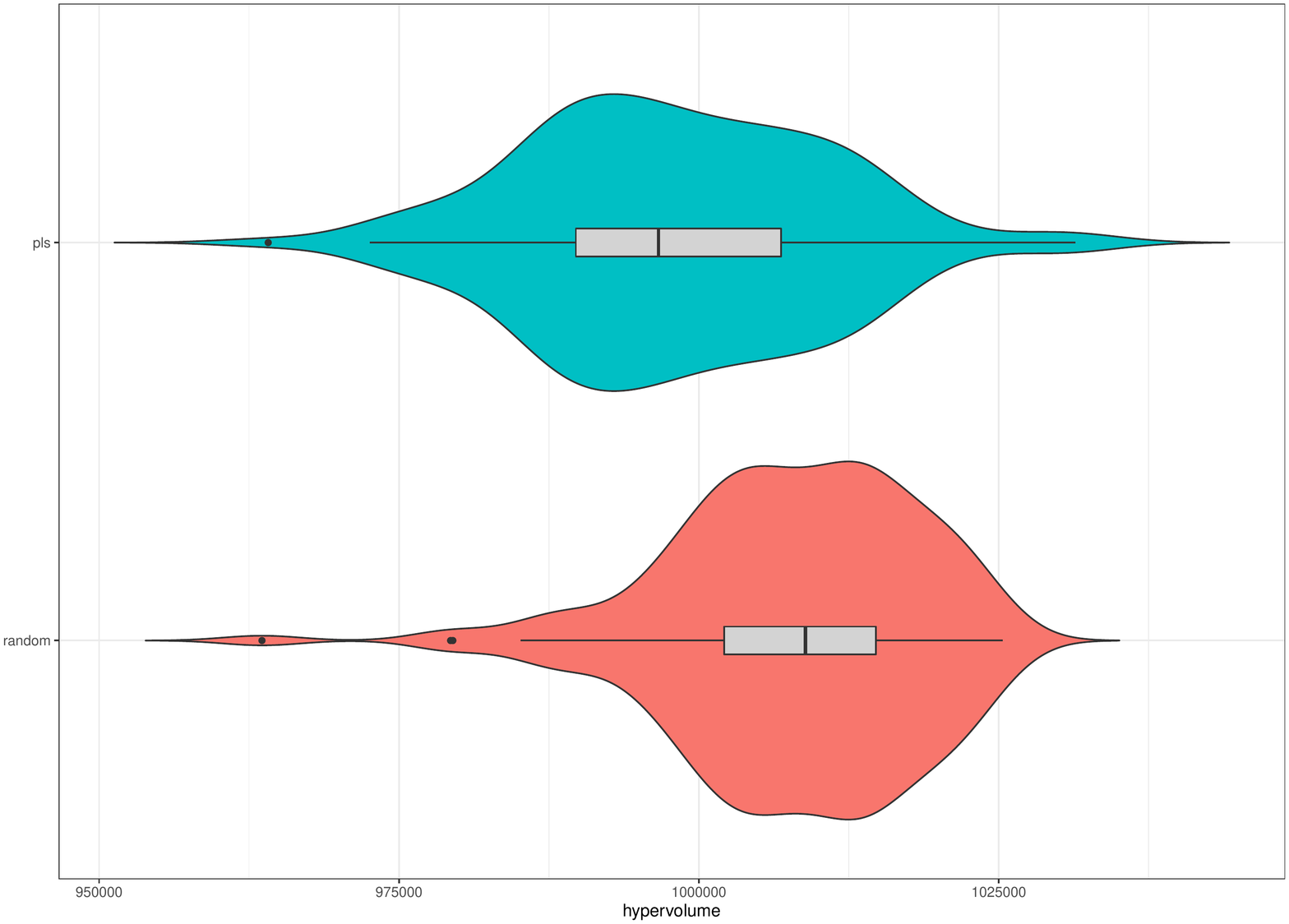}
         \caption{Hypervolume 778 effort}
         \label{fig:nrp100_778_hypervol}
     \end{subfigure}
     \hfill
     \begin{subfigure}[b]{0.47\textwidth}
         \centering
         \includegraphics[width=\textwidth]{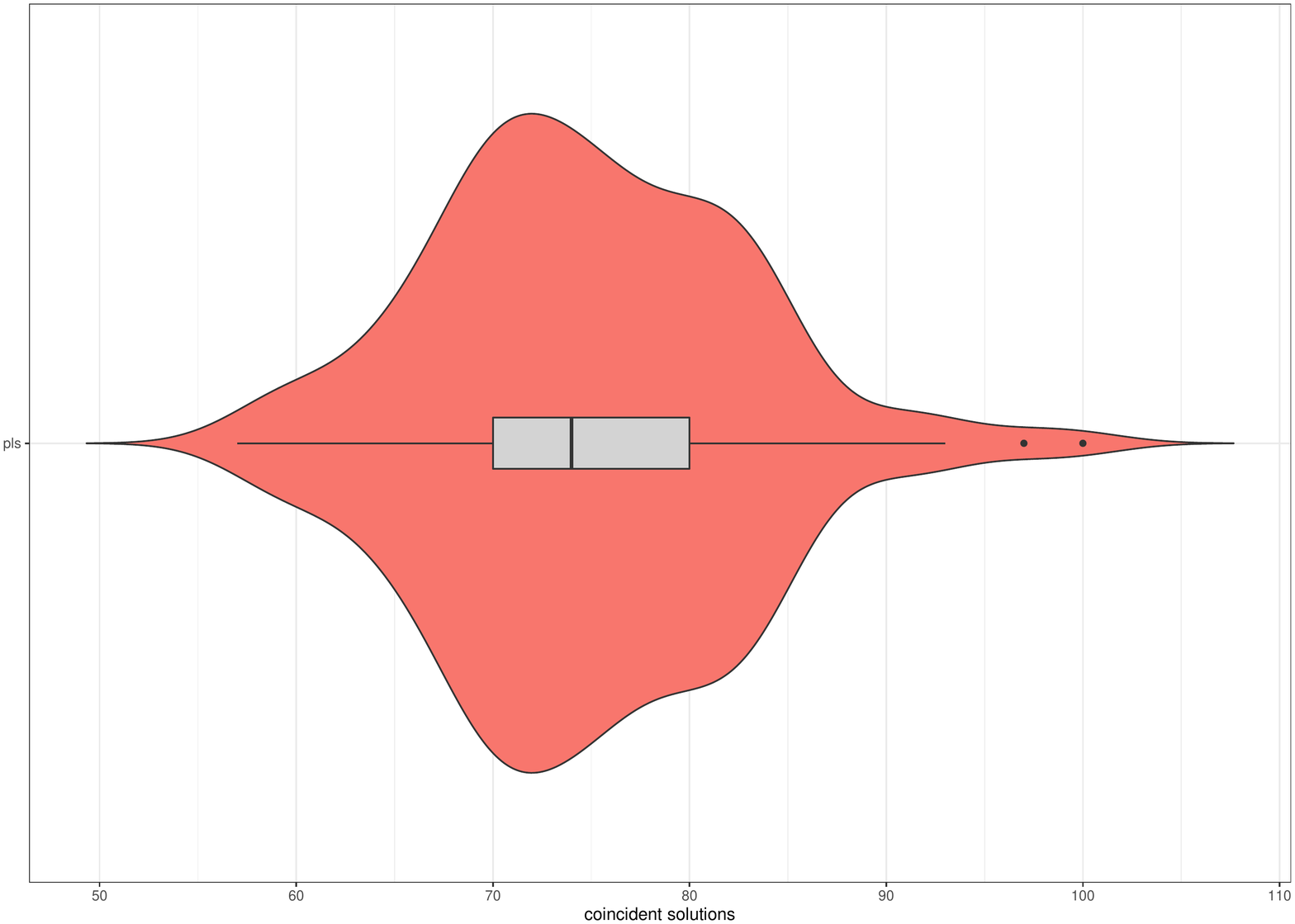}
         \caption{Number of coincident solutions 778 effort}
         \label{fig:nrp100_778_coincident_sols}
     \end{subfigure}
        \caption{NRP100 comparison of the EDA's distributions of hypervolume and number of solutions found that lie in the reference front.}
        \label{fig:nrp100_distributions}
\end{figure}

\subsubsection{NRP with 100 requirements}
In this case, we set three NRP instances corresponding to effort limits of 311, 519 and 778 (a 30\%,  50\% and 75\% of the total problem effort, respectively). Tables \ref{tab:hypervolume} and \ref{tab:coincident_solutions} (see the \emph{exact} column) show, respectively for all instances, the hypervolume and the number of solutions of the reference Pareto front was obtained  dividing ad hoc each instance into four sub-problems (see step 1 in subsection \ref{subsec:methodology}). No time was set because both exact algorithms were unable return a solution within a reasonable time gap (i.e. an hour), this is why we don’t set times necessary to solve the instances. Therefore, We execute the branch and bound algorithm without time limit to get the complete Pareto front for all NRP instances. 

Figure \ref{fig:nrp100_distributions} shows the distributions of hypervolume and number of solutions found that are included in the reference front. As in the previous NRP50 case, and for the same reason (i.e. massive execution times), we discarded the use of  \emph{maxprob} initialization. In all instances, the values of the coefficient of variation for hypervolume are less than 5\% (see column \textit{cv} in Table \ref{tab:hypervolume}), then EDA results can be considered stable (see mean and standard deviation values showed in columns \emph{mean} and \emph{sd} in Table \ref{tab:hypervolume}). However, this is not the case for the number of coincident solutions, where stable results (see mean and standard deviation values showed in columns \emph{mean} and \emph{sd} in Table \ref{tab:coincident_solutions}) are only obtained for the NRP100 instances with effort limits 311 and 519 using \emph{pls} initialization (see column \textit{cv} in Table \ref{tab:coincident_solutions}). 

The Wilcoxon test (see Table \ref{tab:wilcoxon_NRP50_NRP100}) detects significant differences (all $p < 0.05$) between these distributions. In all NRP100 instances, when using \emph{random} initialization better hypervolume values are obtained on average, but the average number of solutions residing on the Pareto front is low (i.e. $90.36$) when the effort limit is $311$ and extremely low for the other two limits (i.e. $1.744$ and $0.248$ for the limits 519 and 778, respectively). In other words, the fronts returned by the EDA using \emph{random} initialisation have a good extension but lie below the reference front in two of the instances. However, for EDA with \emph{pls} initialization, the opposite is true. It obtains, on average, worse hypervolume values and better values in the number of solutions matching those of the reference front. That is, the pareto fronts EDA, with \emph{pls} initialization, returns have less extent, but part of its solutions are in the reference front. Thus, when the effort limit is 311, the average percentage of solutions in the fronts obtained by the EDA, with \emph{pls} initialization, that are in the reference front is 81.01 \%, which represents, also on average, 53.48\% of the total number of solutions in the reference front. For the 519 effort limit these percentages on average are 78.39\% and 45.48\%, respectively, while for the 778 limit they are 32.99\% and 13.80\%.  

With respect to the execution time of the EDA algorithms (see Table \ref{tab:time}) it is worth to note that the number of iterations used was incremented from 100, on the previous NRP instances, to 500. This decision was made mainly due to the number of requirements in NRP100, five times higher than NRP20, with implies a larger search space. The number of iterations has a profound impact on the execution time of EDA algorithms, increasing it, but keeping it high and at a constant level in all cases. The pareto fronts returned by EDA, with \emph{ pls} initialization, have less extent, but part of its solutions are in the reference front. So, the answer to \textbf{RQ3} is still favourable in these large size NRP instance: EDAs are able to find partial approximations to the Pareto fronts, at least if they have time.

\subsubsection{All NRP instances}
We are interested in analysing the performance of EDAs in all NRP instances of different size and complexity in which they were applied. For this, we need performance measures that are independent, reliable and commensurable, such as the average percentage of hypervolume and of the number of coincident solutions. The comparison of these measures for the EDAs is shown in Table \ref{tab:percentages_hypervol_coincident_sols}. Thus, we are faced with two situations. In one, we can compare the performance of the three algorithms (i.e. \emph{random}, \emph{pls} and \emph{maxprob}) but only on three instances of the NRP. While in the other, we can compare two algorithms (i.e. \emph{random} and \emph{pls}) on all NRP instances.  

The comparison of EDA performance with the three initialisation methods \emph{random}, \emph{pls} and \emph{maxprob} is done on three instances of the \emph{nrp20}. Both in percentage of hypervolume and number of coincident solutions, the \emph{p}-values (0.108 and 0.08877, respectively) returned by the Friedman aligned ranks test does not detect statistically significant differences in performance between the three algorithms.

In the case of \emph{random} and \emph{pls} initialisations, the performance analysis can be extended to all instances of the \emph{nrp}. However, the Wilcoxon signed-rank test also does not detect statistically significant differences in either the percentage of hypervolume (\emph{p}-value 0.5703), or that of coincident solutions (\emph{p}-value 0.1641).

%%%%%%%%%%% CONMENSURABLE MEASURES FOR PERFORMANCE
\begin{table}[htpb]
\caption{Comparison of the percentages of hypervolume and coincident solutions for the EDAs}
\label{tab:percentages_hypervol_coincident_sols}
\resizebox{\textwidth}{!}{
\begin{tabular}{lllccclccc}
\hline
\multicolumn{2}{c}{\textbf{Instance} }  &  & \multicolumn{3}{c}{\textbf{\% hypervolumen}} &  & \multicolumn{3}{c}{\textbf{\% coincident sols}} \\ \cline{4-6} \cline{8-10} 
\multicolumn{2}{l}{}            &  & random & pls    & maxprob &  & random & pls    & maxprob \\ \cline{1-2} \cline{4-6} \cline{8-10} 
       & \multicolumn{1}{r}{25} &  & 0.9993 & 0.9999 & 0.9999  &  & 0.9642 & 0.9663 & 0.9958  \\
nrp20 & \multicolumn{1}{r}{43} &  & 0.9820     & 0.9890     & 0.9986    &  & 0.8548      & 0.8815      & 0.9467     \\
       & \multicolumn{1}{r}{60} &  & 0.9688 & 0.9802 & 0.9942  &  & 0.7375 & 0.7738 & 0.8600  \\ \cline{1-2} \cline{4-6} \cline{8-10} 
       & 107                    &  & 0.9993 & 0.9956 &         &  & 0.9606 & 0.8348 &         \\
nrp50  & 179                    &  & 0.9904 & 0.9970 &         &  & 0.7607 & 0.8069 &         \\
       & 268                    &  & 0.9905 & 0.9899 &         &  & 0.7570 & 0.7168 &         \\ \cline{1-2} \cline{4-6} \cline{8-10} 
       & 311                    &  & 0.9755 & 0.8679 &         &  & 0.3530 & 0.5348 &         \\
nrp100 & 519                    &  & 0.9383 & 0.8431 &         &  & 0.0041 & 0.4549 &         \\
       & 778                    &  & 0.8964 & 0.8877 &         &  & 0.0005 & 0.1380 &         \\ \hline
\end{tabular} }
\end{table}

%TBC
\section{Threats to Validity}
The potential threats weakening how trustworthy and generalisable is any research work \cite{wohlin2012} are discussed next, describing also what has been done to mitigate them.

%construct
Construct validity concerns the relation between theory and observation. We use the hypervolume metric and the number of solutions found that are in the reference Pareto front, to assess the quality of the results. Although the hypervolume measures the convergence to the reference front and the spread of the solutions found, the number of coincidences complete this information in the case of the approximate algorithms proposed.

%internal and conclusion
The internal validity threat is related to the applied experimental methodology and the causality relationships that are being examined. In our study, there is inherent stochasticity both in execution times and results. To mitigate this randomness, we have performed several executions for every NRP instance and statistical procedures to evaluate the results returned. Specifically, for the analysis of the independent executions of the algorithms within each NRP instance we used non-parametric tests: Wilcoxon test with Holm's correction when comparing two algorithms or the Kruskal-Wallis test with a posthoc analysis using the Connover-Iman test when three algorithms were involved. Whereas the comparison of the performance of the algorithms over all NRP instances has been carried out using the average percentage of the hypervolume and of the number of matching solutions. In this case, we used the Wilconxon signed-rank test when comparing two algorithms or Friedman aligned rank test combined with a posthoc analysis using the Friedman test with Finner's correction to check the differences between pairs, when comparing three algorithms. Thus, results (i.e. conclusion validity threat) have been interpreted by taking into consideration statistical tests in order to mitigate the potential errors caused by stochasticity.

%external
Finally, regarding external validity (or the capability to generalize our results), we use different data sets to analyse the generalization of the observed results. Each of them varies in number of requirements and constraints, and correspond to problems that have been used previously, in the NRP domain, to test algorithms. These problems have been studied using three different effort bounds obtaining nine instances of the NRP.
%Each of them varies in number of requirements and constraints and corresponds toproblem that have been previously used in the field of the NRP problem to test algorithms have been studied using three different effort limits so as to use nine instances of NRP.
%\color{magenta} Unfortunately as far as we know, there is a lack of real datasets due to the privacy policies followed by software development companies. Hence, the impossibility of using other data sets linked to real applications.\color{black}

\section{Conclusions}

The bi-objective Next Release Problem, as many other problems in Software Engineering, can be modelled as a multi-objective optimization problem. In it,  requirements interactions play a crucial role as they have to be fulfilled at the time of assembling requirements to form a solution. We embedded interactions into an exact (i.e. branch and bound algorithm) and an approximate algorithm (i.e. EDA) to find a set of non-dominated solutions for a NRP instance. The use of an interaction graph do enhance the search, but the enhancement achieved is not enough to discard the use approximate methods (specially if time restrictions exists, as in the NRP case). This graph provides a visual explanation of requirements interactions and serves as a guideline for adding requirements to form valid solutions, as the branch and bound algorithm has shown, and  it defines the structure of the probabilistic graphical model which is the core of the proposed EDA algorithm.
In addition, evolutionary algorithms could also take advantage of the interaction graph when generating the populations, as it allows the construction of solutions that satisfy the conditions of the problem and avoids the additional effort of constructing them and checking whether they are valid or not.

In practice, EDAs have proven to be stable, reliable and fast, at least in small and medium size NRP cases. For the different instances of the problem, in few seconds, they have obtained a set of well-spread solutions, many of them located on the reference front. It is worth to notice that, when using the EDA, there are often multiple possible node ordering in simulation. However, node ordering has no impact on the sampling distribution for a particular node: the distribution is always based on the node's predecessors, which are always instantiated before the node itself is. 
Different population initialization methods have been tried (i.e. \emph{random}, \emph{pls} and \emph{maxprob}), but, with the exception of \emph{maxprob} which was discarded due to its high execution times, neither \emph{random} nor \emph{pls} can be chosen as the best alternative in small and medium size cases, whereas \emph{pls} has proven to be better in large size cases. With respect to the other parameters, some research remains to be done to obtain clear conclusions about their configuration. 

The research we have carried out provides answers to the research questions posed. Thus, for the \emph{\textbf{RQ1} How can requirements interactions be included in search algorithms to find the Pareto front of a next release problem?} we have shown how a branch and bound algorithm can incorporate requirements interactions into the search process and explore possible solutions by inserting requirements following that order. Likewise, we have defined an EDA which uses a probabilistic graphical model based on the fixed structure defined by requirements interaction (i.e. interaction graph) to obtain populations by sampling it. These are the ways we have explored to incorporate requirements interactions into algorithms, and that give the answer to \emph{\textbf{RQ1}}.

To answer question \emph{\textbf{RQ2} Is there any improvement in incorporating the interactions between the requirements in the algorithms?}, we can observe the percentages of relative change in the instances of NRP20 and NRP50. They indicate there is a reduction in execution time when requirements interactions are included in the search process (i.e. branch and bound algorithm), with respect to that of the brute force algorithm. Even in the case of our EDA there is an improvement in the execution times for these instances. Besides, in the instances of the large size NRP100, when no time limit is set, at least EDA can give a partial approximated solution within reduced time. This gives rise to the last question: \emph{\textbf{RQ3} If a set of non-dominated solutions is required in a short time, do estimation of distribution algorithms find a high quality (i.e. well-spread) one?} For small and medium size NRP cases, NRP20 and NRP50, EDAs with random and pls initialization are able to find good approximations to the Pareto fronts in a limited time. But in the large size NRP case, NRP100, EDAs are able to find fronts that are partial approximations of the reference ones, if they have time. In our experiments we found no statistically significant differences in the performance of the EDAs. However, in large instances of the NRP100, the pls initialization locates a larger number of solutions that match those of the reference fronts but at the cost of losing extension in the fronts it provides.

In summary, we have shown what is the requirements interactions impact when searching for solutions of the bi-objective Next Release Problem. We have explicitly included the interactions of the requirements in an exact (i.e. branch and bound) and an approximate algorithm (i.e. estimation of distribution algorithm). And we found that interactions inclusion do enhance the search and when time restrictions exists, as in the case of small and medium size cases of the bi-objective Next Release Problem. In them, EDAs have proven to be stable and reliable locating a large number of solutions on the reference Pareto front. However, for large size cases, EDAs returned partial approximations. Consequently, parameter settings and ways to speed up the calculation of EDA remain to be investigated. \color{black} 

\section*{Acknowledgements}
\noindent
This research has been funded by the Spanish Ministry of Science, Innovation and Universities under the Search based software engineering research network (RED2018-102472-T), and under project PID2019-106758GB-C32 (EML-PA). It is also partially supported by Data, Knowledge and Software Engineering (DKSE) research group (TIC-181) of the University of Almer\'{\i}a. We would also like to thank the anonymous reviewers for their useful comments, which have helped to improve the quality of the paper.

%% Loading bibliography style file
% \bibliographystyle{model1-num-names}

% Loading bibliography database

\end{document}